\documentclass[12pt,preprint]{aastex}

\begin{document}
\title{Characteristics of Gamma-Ray Loud Blazars in the VLBA Imaging and Polarimetry Survey}
\author{J. D. Linford\altaffilmark{1}, G. B. Taylor\altaffilmark{1}, R. Romani\altaffilmark{2}, S. Healey\altaffilmark{2}, J. F. Helmboldt\altaffilmark{3},  A. C. S. Readhead\altaffilmark{4}, R. Reeves\altaffilmark{4}, J. Richards\altaffilmark{4} and G. Cotter\altaffilmark{5}}

\altaffiltext{1}{Department of Physics and Astronomy, University of New Mexico, MSC07 4220, 1 University of New Mexico, Albuquerque, NM 87131-0001, USA}
\altaffiltext{2}{Department of Physics, Stanford University, Stanford, CA 94305}
\altaffiltext{3}{Naval Research Laboratory, Code 7213, 4555 Overlook Ave. SW, Washington, DC 20375}
\altaffiltext{4}{Astronomy Department, California Institute of Technology, Mail Code 247-17, 1200 East California Boulevard, Pasadena, CA 91125}
\altaffiltext{5}{Department of Astrophysics, University of Oxford, Denys Wilkinson Building, Keble Road, Oxford OX1 3RH, United Kingdom}
\received{?}
\accepted{?}

\begin{abstract}

The radio properties of blazars detected by the Large Area Telescope (LAT) on board the \textit{Fermi Gamma-ray Space Telescope} have been observed as
part of the VLBA Imaging and Polarimetry Survey (VIPS).  This large, flux-limited sample of active galactic nuclei (AGN) provides insights into the mechanism that produces strong $\gamma$-ray emission.  At lower flux levels, radio flux density does not directly correlate with $\gamma$-ray flux.  We find that the
LAT-detected BL Lacs tend to be similar to the non-LAT BL Lacs, but 
that the LAT-detected FSRQs are often significantly different from the non-LAT FSRQs.  
The differences between the $\gamma$-ray loud and quiet FSRQs can be explained
by Doppler boosting; these objects appear to require larger Doppler factors
than those of the BL Lacs.  It is possible that the $\gamma$-ray loud FSRQs are fundamentally different from the $\gamma$-ray quiet FSRQs.  Strong polarization at the base of the jet appears to be a signature for $\gamma$-ray loud AGN.

\end{abstract}


\section{Introduction}

Several outstanding questions about $\gamma$-ray loud active galactic nuclei (AGN) remain, including: ``why are some sources $\gamma$-ray loud and others are $\gamma$-ray quiet?''  The answer to this is likely related to Doppler boosting (Jorstad et al. 2001; Taylor et al.\ 2007; Kovalev et al.\ 2009; Lister et al.\ 2009a; Savolainen 2010), though many details remain to be filled in.  A related question is ``where are the $\gamma$-rays produced: in or near the core or throughout the jets?''  There are arguments for both locations (Sikora et al.\ 2009; Tavecchio et al.\ 2010; Marscher et al.\ 2006, 2008, 2010).  In either case it still is unclear how the radio flaring, or the ejection of new components, is related to the gamma-ray activity, or
whether there are multiple classes of $\gamma$-ray blazars (e.g., Sikora et al.\ 2001, 2002).  In addition, Stawarz et al.\ (2008) argue that we should detect $\gamma$-rays from the compact ($<$kpc) lobes of young radio sources.

The Large Area Telescope (LAT; Atwood et al.\ 2009) on board the \textit{Fermi Gamma-ray Space Telescope} is a wide-field telescope covering the energy range from about 20 MeV to more than 300 GeV.  At high galactic latitudes 80\% (106 of 132) of the $\gamma$-ray
bright sources detected in the LAT Bright Source List (BSL) derived
from the first 3 months of \textit{Fermi} observations (Abdo et al.\ 2009a) are
associated with known Active Galactic Nuclei (AGN).  This trend
continues in the LAT first-year catalog (1FGL; Abdo et al.\ 2010a),
where 662 of 1043 high latitude sources are associated with AGN with
high confidence ($P >$ 80\%).  The vast majority of the LAT 
$\gamma$-ray sources are blazars, with strong, compact radio emission.
These blazars exhibit flat radio spectra, rapid variability, compact
cores with one-sided parsec-scale jets, and superluminal motion in the
jets (Marscher 2006).  A handful of radio galaxies and Seyferts have
also been identified in the First LAT AGN Catalog (1LAC; Abdo et al.\ 2010b).  These radio galaxies (e.g., 3C84,
M87, Cen A) have faint parsec-scale counterjets indicating that they
are viewed at larger angles to the line-of-sight than the blazars.
Also, there is a much larger fraction of unclassified sources in the 1
year catalog than was present in the 3 month catalog.

The VLBA Imaging and Polarimetry Survey (VIPS; Helmboldt et al.\ 2007) was originally devised to obtain VLBA measurements of AGN in the fifth data release of the Sloan Digital Sky Survey (DR5; Adelman-McCarthy et al.\ 2006).  It consists of 5 GHz VLBA images and polarization data of 1127 AGN located between declinations of $15^\circ$ and $65^\circ$.  The VIPS sources were drawn from the Cosmic Lens All-Sky Survey (CLASS; Myers et al.\ 2003).  All VIPS sources were required to have CLASS flux densities of 85 mJy or more.  The VLBA at 5 GHz gives us an angular resolution of just under 2 milliarcseconds, which allows us to probe sub-parsec scale features of the objects.  Using multi-epoch observations, we will be able to measure the apparent jet velocities.

Various programs, such as MOJAVE observing at 15 GHz (Lister et al.\ 2009a; Homan et al.\ 2009), the Boston University program 
observing at 22 and 43 GHz (Roca-Sogorb et al.\ 2010), and TANAMI (Ojha et al.\ 2010) observing at 8.4 and 22 GHz, are in place to
monitor the radio jets from the brightest blazars such as 3C273, BL
Lac, etc.  The detection of Centaurus~A, 3C84 and M87 by \textit{Fermi} (Abdo et
al.\ 2009a, 2009b, 2009c) and the detection of M87 at TeV energies (Aharonian et
al.\ 2006) suggests that other classes of AGN (e.g., radio galaxies)
may form a significant population of $\gamma$-ray emitters.  Many of the
current programs are strongly biased towards the radio bright sample.
  VIPS has a
flux limit roughly an order of magnitude below the MOJAVE survey and
most other samples that have been used to study the properties of
$\gamma$-ray blazars and so allows us to probe the extension of the radio core/$\gamma$-ray properties down to a fainter population.  Many different radio-$\gamma$-ray correlations have been suggested (Taylor et al.\ 2007, Abdo et al.\ 2009a, Lister et al.\ 2009a, Kovalev et al.\ 2009, Giroletti et al.\ 2010).  By examining a larger sample (1127 objects) we attempt to obtain more definitive insight into the properties of $\gamma$-ray loud blazars.

In \S2 we define our sample.  In \S3 we present data on the $\gamma$-ray and radio properties of the LAT sources. In \S4 we compare several parameters of the LAT and non-LAT sources. Notes on some individual objects of interest are presented in \S5 and in \S6 we discuss our results. Some conclusions are given in \S7.  Throughout this paper we assume $H_{0} = 71$ km s$^{-1}$ Mpc$^{-1}$ and $\Lambda$CDM cosmology (e.g., Hinshaw et al.\ 2009).

\section{Sample Definition}

We have compared the 11 month First LAT AGN Catalog (1LAC) to the
VLBA Imaging and Polarimetry Survey (VIPS; Helmboldt et al.\ 2007), and from the intersection compiled a
sample of 102 LAT sources at $|b| > 10^\circ$ associated with an AGN with a probability of more than 50\% (although most of the probabilities are greater than 98\%).  All of these objects were observed with the Very Long Baseline Array (VLBA) from 1996 to 2006 (although 958 of the 1127 sources were observed during 2006) at 5 GHz.  Each VIPS source was observed for approximately 500 seconds, with scans spread out over 11 hours.  See Table~\ref{datatable} for a summary of our data.  Of the LAT objects, 40 are BL Lacs, 50 are FSRQs, and 12 are either radio galaxies or unclassified objects.  The optical classification were adopted from the Candidate Gamma-Ray Blazar Survey (CGRaBS; Healey et al. 2008), except that objects labeled in CGRaBS as `AGN' we call radio galaxies.

To build a $\gamma$-ray quiet sample for comparison, we excluded all LAT sources from the VIPS sample, leaving 1018 objects in our non-LAT sample.  Of these 1018 objects, 24 are BL Lacs, 479 are FSRQs, and 515 are either radio galaxies or unclassified objects.

\section{Gamma-ray Flux and Radio Flux Density}

The radio flux densities in this section, and those listed in Table~\ref{datatable}, are measured at 8.5 GHz from CLASS (Myers et al.\ 2003).  For this section we also include 7 MOJAVE-detected $\gamma$-ray loud sources for which we had CLASS flux densities, increasing the size of our LAT sample to 109 sources.

\subsection{Redshift Selection Effect}

Throughout the rest of this paper, we will use the nonparametric Spearman test (e.g., Press et al.\ 1986) to look for correlations between the LAT-detected and non-LAT-detected objects.  The Spearman correlation coefficient ($\rho_{s}$) has a range of $0<|\rho_{s}|<1$.  A high value of $\rho_{s}$ indicates a significant correlation.  This is a powerful test for statistical correlation, but it does not test an actual physical correlation.  In order to be certain our correlations are physically significant, we make sure that a redshift selection effect is not adding a bias to our data.  

Imagine a population of sources covering a wide range of redshifts in which the radio and $\gamma$-ray emission is not physically correlated, but in which there is a correlation of radio flux, and gamma-ray flux, with redshift.  This is actually what one would naively expect because the more distant objects will be fainter.  Such a population will show a strong correlation of radio flux with gamma ray flux via the Spearman test, but this does not indicate a significant physical correlation between these two observables.  

\begin{figure}
\plotone{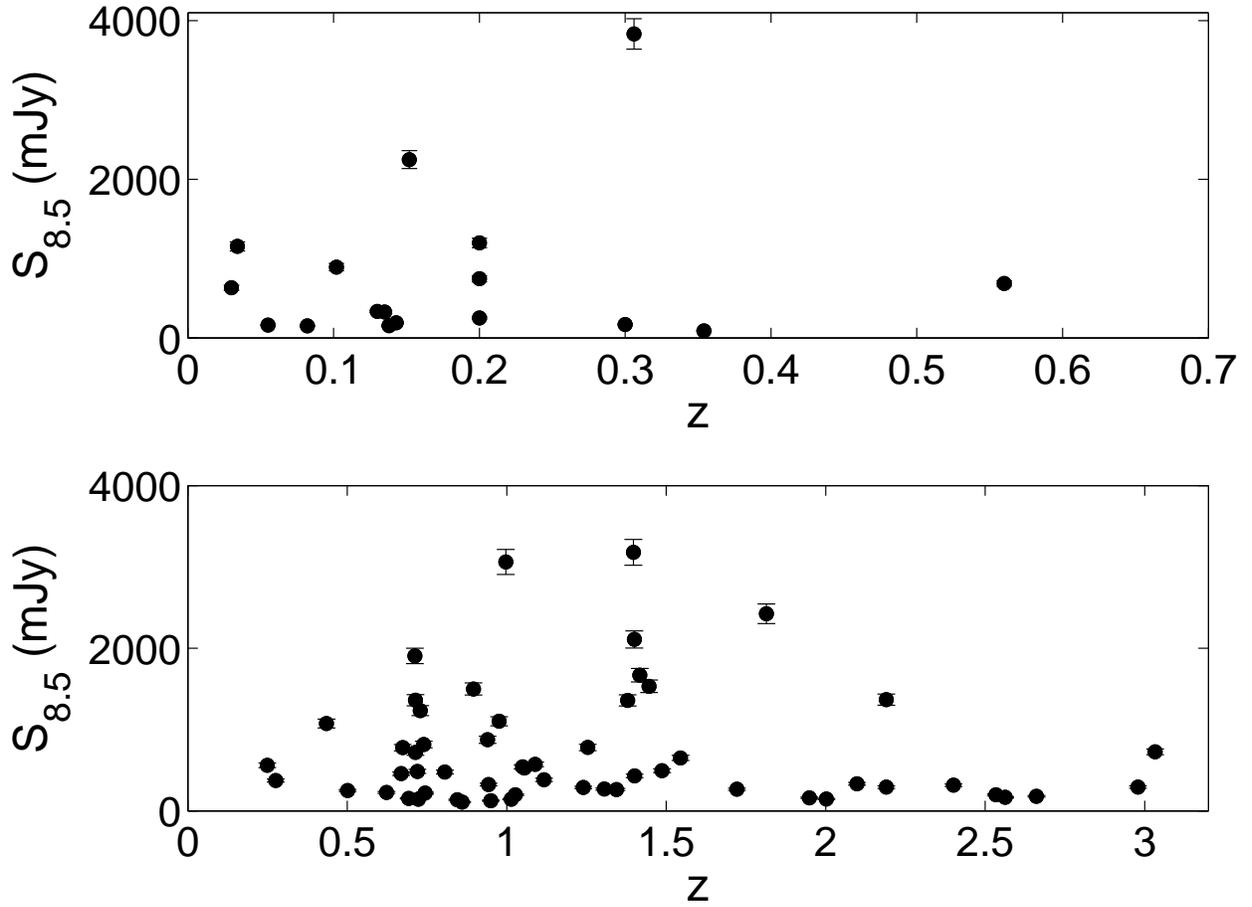}
\caption{CLASS 8.5 GHz radio flux density (in mJy) versus redshift of LAT-detected sources. BL Lacs are the top plot, FSRQs are the bottom plot.}
\label{rvz}
\end{figure}

However, by investigating the relationship between radio flux density and redshift for the LAT-detected sources (see Fig.~\ref{rvz}) we can rule out a correlation between the two.  We calculated the Spearman $\rho_{s}$ values for the $S_{8.5}$-z relationship for the BL Lacs and FSRQs separately.  For the BL Lacs, the $\rho_{s}$ value is 0.0737, with a 79\% probability that random sampling would produce this same $\rho_{s}$ value.  For the FSRQs, the $\rho_{s}$ value is 0.0450, with a 74\% probability of getting the same value by random sampling.  Therefore, there is no significant correlation between the radio flux density and redshift for our objects.

\begin{figure}
\plotone{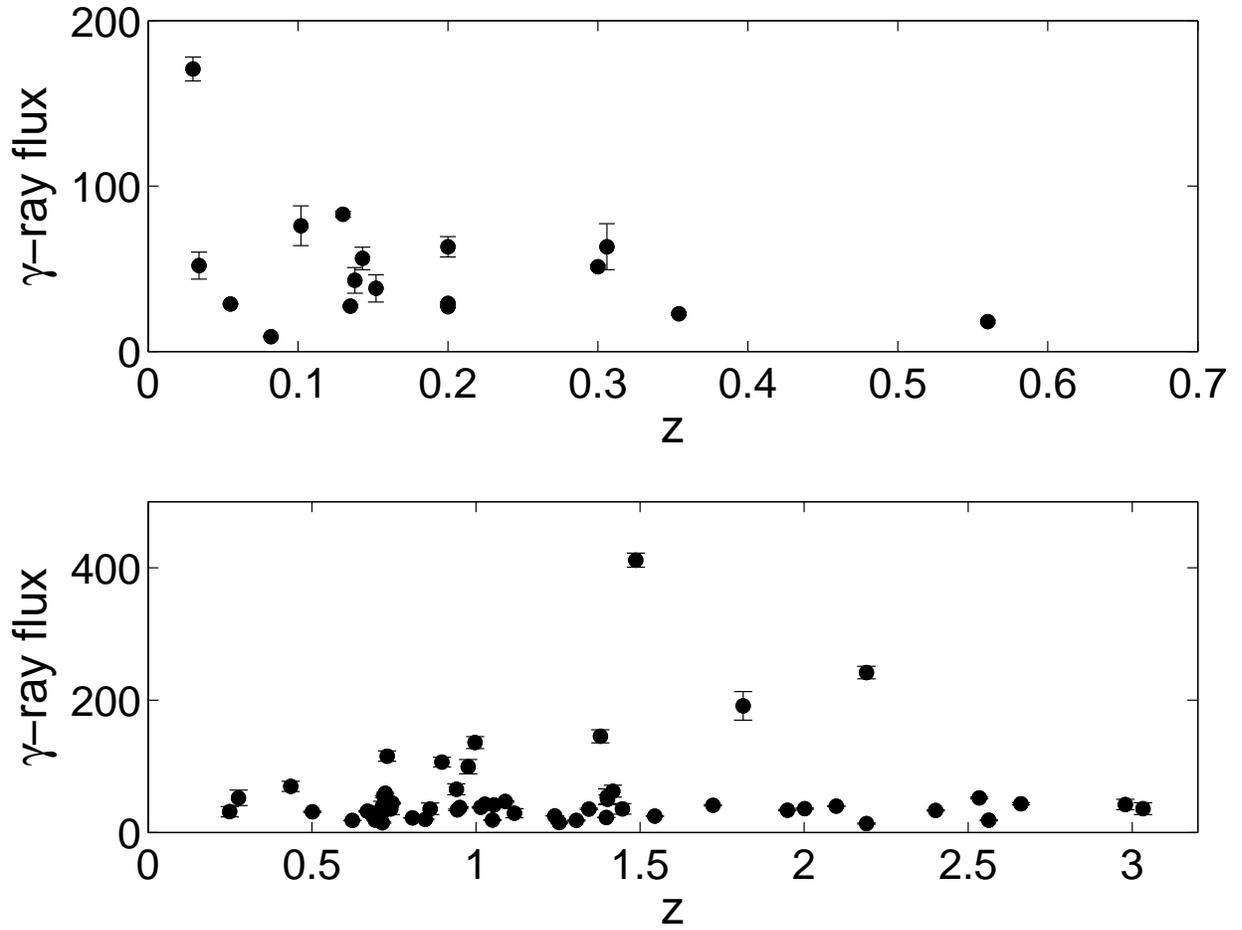}
\caption{LAT $\gamma$-ray flux (100 MeV - 100 GeV) versus redshift.  The $\gamma$-ray fluxes are in units of  $10^{-9}$ photons cm$^{-2}$ s$^{-1}$.  BL Lacs are the top plot, FSRQs are the bottom plot.}
\label{gvz}
\end{figure}

For completeness, we also tested the correlation between $\gamma$-ray flux and redshift.  The $\rho_{s}$ values are -0.319 for BL Lacs and 0.104 for FSRQs, with the probability of getting the same values from random sampling of 21\% and 45\%, respectively.  So, there is no significant correlation between $\gamma$-ray flux and redshift, either.  See Fig.~\ref{gvz} for a plot of $\gamma$-ray flux versus redshift.

\subsection{Gamma-ray Flux vs. Radio Flux Density}

\begin{figure}
\plotone{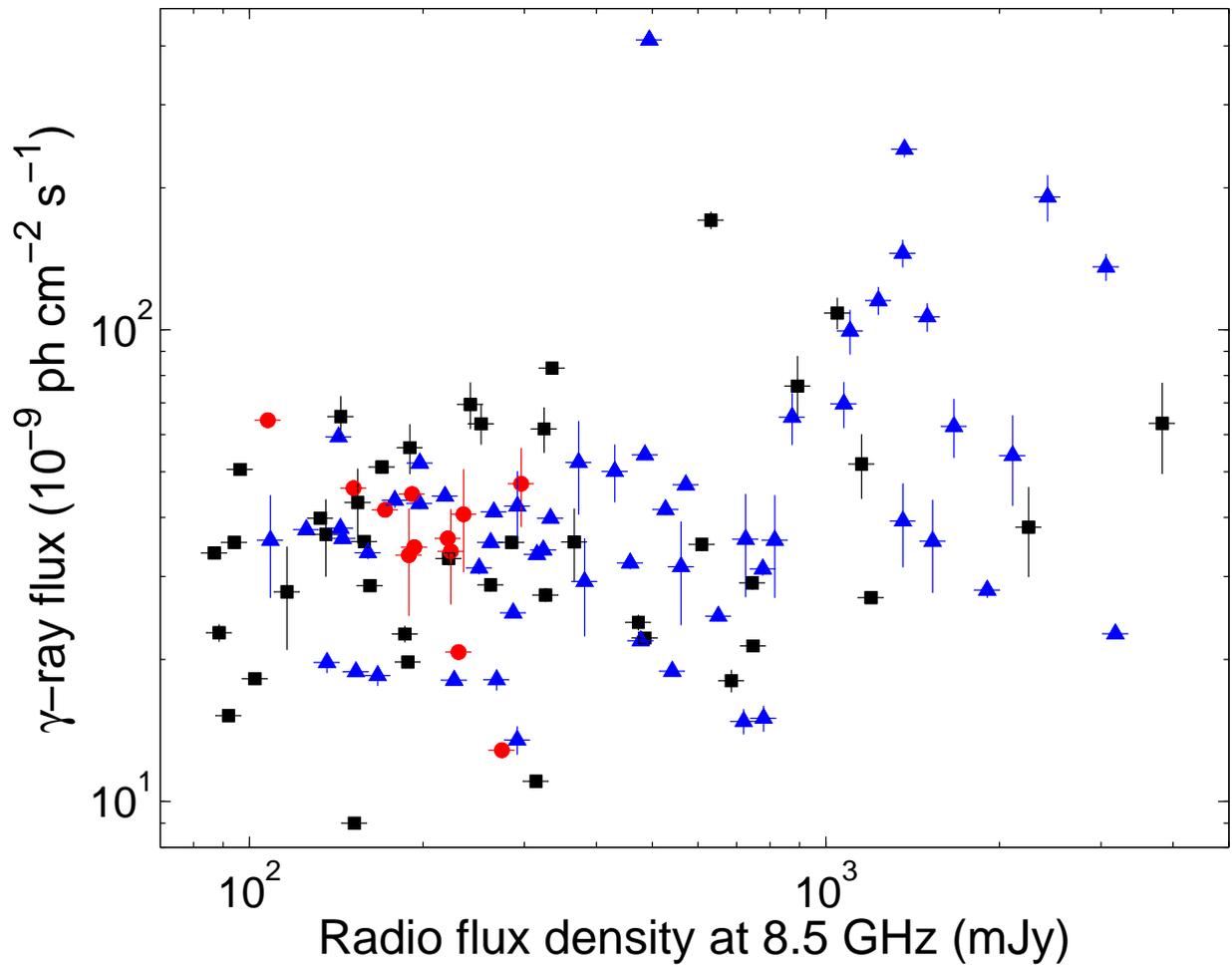}
\caption{LAT $\gamma$-ray flux (100 MeV - 100 GeV) versus CLASS radio flux density at 8.5 GHz.  The $\gamma$-ray fluxes are in units of $10^{-9}$ photons cm$^{-2}$ s$^{-1}$.  The black squares are BL Lacs, the blue triangles are FSRQs, and the red circles are radio galaxies and unclassified objects.}
\label{FluxFlux}
\end{figure}

In Fig.~\ref{FluxFlux} we plot the LAT flux versus the flux density at 8.5 GHz from the CLASS.  The LAT fluxes are broadband fluxes from 100 MeV to 100 GeV.  The BL Lacs have a large range in their flux density and $\gamma$-ray flux, but one can see a large fraction of them have relatively low radio flux density ($S_{8.5}<$ 400 mJy) but high $\gamma$-ray flux.  The FSRQ objects also have a very large range in both their radio flux density and their $\gamma$-ray flux, but the higher flux density FSRQs also tend to have higher $\gamma$-ray flux.  
Overall, there is a lack of strong $\gamma$-ray emitters among the weak radio sources, but the weak $\gamma$-ray emitters can be bright or faint at radio wavelengths.  
The median $\gamma$-ray flux is $3.5 \times 10^{-8}$ photons cm$^{-2}$ s$^{-1}$ for BL Lacs, $3.7 \times 10^{-8}$ photons cm$^{-2}$ s$^{-1}$ for FSRQs, and $3.8 \times 10^{-8}$ photons cm$^{-2}$ s$^{-1}$ for the radio galaxies and unclassified objects.

The Spearman $\rho_{s}$ values we found were 0.217 for the BL Lacs, 0.318 for the FSRQs, and -0.413 for the radio galaxies and unclassified objects.  For the BL Lacs and radio galaxies/unclassified objects, the chances that random samplings would result in the same $\rho_{s}$ values were greater than 16\%, indicating a lack of correlation between radio flux density and $\gamma$-ray flux.  For the FSRQs, the chance of a random sampling having the same $\rho_{s}$ was only 1.7\%, which indicates that there is a significant correlation between 8.5 GHz flux density and $\gamma$-ray flux for the FSRQs.  However, when we exclude the brightest 10\% of sources in the radio, we find that the remaining FSRQs no longer correlate.  The weaker 90\% of sources have $\rho_{s}$'s of 0.135 for BL Lacs and 0.159 for FSRQs, with the chances of random samples having the same $\rho_{s}$ values of 18\% for BL Lacs and 33\% for FSRQs.  

The lack of correlation for BL Lacs is counter to what has been reported in other studies.  Kovalev et al.\ (2009) find a definite correlation for their MOJAVE sources at 15 GHz.  Their radio data was taken closer to the LAT measurements, and the $\gamma$-ray blazars are suspected to have higher radio flux densities during or shortly after $\gamma$-ray emission (Pushkarev et al. 2010).  However, their sample is only about 1/3 the size of ours and it is biased to sources with high radio flux density ($>$1.5 Jy).  Very strong flux-flux correlations are found by Ghirlanda et al.\ (2010) at 20 GHz.  They have a larger sample than ours with 230 LAT sources, about half of which are FSRQs.  However, their sample is also mostly comprised of bright sources.  More than half of their sources have $S_{20}>500$ mJy, compared to our sample where about 1/3 have $S_{8.5}>500$ mJy.

To see if we could reproduce the correlation seen by others, we applied the Spearman test to the 36 sources with 8.5 GHz flux densities above 500 mJy.  This subsample is roughly the brightest third of our overall sample and contains 11 BL Lacs and 25 FSRQs.  At first, we did not break the subsample into optical types.  The result is a $\rho_{s}$ of 0.390.  The chance of a random sampling having the same $\rho_{s}$ value as our strong source subsample was only 1.9\%, indicating that this is a significant correlation.  When the subsample was broken up by optical type, the $\rho_{s}$'s were 0.214 for BL Lacs and 0.422 for FSRQs.  The chances of random samples haveing the same $\rho_{s}$'s were 69\% and 3.7\%, respectively.  So we find that FSRQs exhibit a significant correlation between radio flux density and $\gamma$-ray flux but BL Lacs do not, even when limited to the strongest radio sources.  Any sample that contains mostly high radio flux density objects will automatically be dominated by FSRQs and will likely show a correlation between radio flux density and $\gamma$-ray flux.  

\subsection{Radio Flux Density}

The median flux density at 8.5 GHz is 247 mJy for the LAT BL Lacs and 276 mJy for the non-LAT BL Lacs.  The LAT FSRQ appear to have significantly higher flux densities than the non-LAT FSRQ, with a median of 377 mJy for LAT FSRQs compared to a median of 238 mJy for non-LAT FSRQs.  The LAT radio galaxies and unclassified objects also have higher 8.5 GHz flux densities than the non-LAT sources.  The median flux density for LAT radio galaxies/unclassified objects is 207 mJy and the median flux density for non-LAT radio galaxies/unclassified objects is 119 mJy.  The K-S test probability that the LAT and non-LAT BL Lac objects belong to the same parent population is about 81\%.  The K-S test probability that the LAT and non-LAT FSRQs are related is only 0.3\%.  The radio galaxy/unclassified objects K-S test probability is 0.01\%.

\section{Morphological Comparison}

\subsection{Source Classes}

VIPS sources are classified as point sources (PS), short jets (SJET), long jets (LJET), complex (CPLX), or compact symmetric object candidates (CSO) by the automatic classification scheme of Helmboldt et al.\ (2007).  See Table~\ref{morphtable} for the source classifications for both LAT and non-LAT sources in VIPS.  From the classifications, it appears that short jet and point source BL Lacs are less likely to produce $\gamma$-rays.  About 75\% of the LAT BL Lacs are classified as LJET, compared to only 45\% for the non-LAT BL Lacs.  The major difference between LAT and non-LAT FSRQs is the lack of CSOs in the LAT sample.  There is also a lack of CSOs among the LAT radio galaxies and objects without an optical classification.

\subsection{Polarization}

\begin{figure}
\plotone{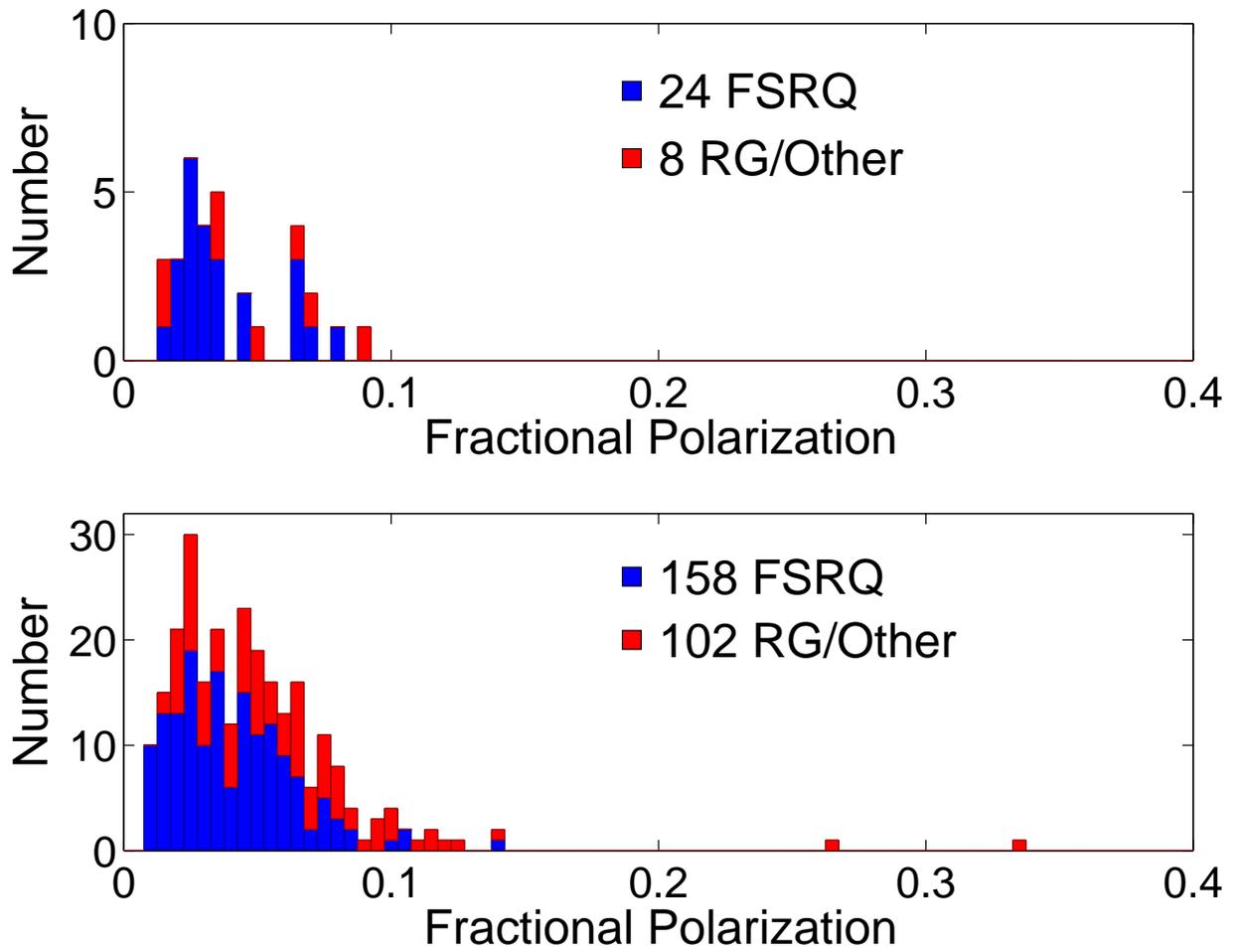}
\caption{The distributions of fractional polarization in the cores of LAT-detected FSRQs and radio galaxies/unclassified objects (top) and non-LAT FSRQs and radio galaxies/unclassified objects (bottom).}
\label{polFSRQnO}
\end{figure}

Polarization is detected from the cores of nearly half (49 of 102) of the LAT sources in our sample.  We define the core as the bright compact component at the base of the jet.  We detected polarization in the core of 17 of the 40 BL Lacs, 24 of the 50 FSRQs, and 8 of the 12 radio galaxies and unclassified objects.  For the non-LAT sources, core polarization is detected in only 270 of the total 1018 sources.  There are 10 of 24 non-LAT BL Lacs, 158 of 479 FSRQs, and 102 of 515 radio galaxies or unclassified objects detected.  The LAT FSRQs and radio galaxies/unclassified objects tend to have polarization in their cores more than their non-LAT counterparts.  In contrast, the LAT BL Lacs do not appear to be any more likely to have polarized cores than the non-LAT BL Lacs.  The distributions of fractional polarization in the cores for LAT and non-LAT FSRQs, and radio galaxies/unclassified objects are shown in Fig.~\ref{polFSRQnO}.  
The Kolomogorov-Smirnov (K-S) test probabilities that these distributions were drawn from the same parent populations are 84.6\% for the BL Lacs, 39.4\% for the FSRQs, and 76.9\% for the radio galaxies/unclassified objects.  

Our polarization results are different than those reported by Hovatta et al. (2010) with their MOJAVE data at 15 GHz.  They found that the median fractional polarization for the cores of LAT-detected sources were higher than for non-LAT sources.  Their median values for their 2008-2009 observations are 2.5\% for LAT-detected sources and 1.86\% for non-LAT sources.  Our overall median values are 3.5\% for LAT sources and 4.4\% for non-LAT sources.  However, our result is not of high significance as the standard deviations are 2.5\% for LAT sources and 3.4\% for non-LAT sources.  It should also be noted that our VIPS data is not contemporaneous with LAT observations.  It is possible that polarization increases with $\gamma$-ray emission.  In fact, the MOJAVE data indicates that the polarization levels were stronger during the LAT detections than in previous years (2002 - 2008), but only by a factor of 1.2.

\subsection{Core Brightness Temperature}

We have obtained the brightness temperatures from automatic modelfits
to source components as described in Helmboldt et al. (2007).  
For sources with two or fewer components we have further refined 
our modelfitting procedure 
by fitting to the visibility data directly (Taylor et al.\ 2007).  Only image plane 
modelfitting was carried out in Helmboldt et al.\ (2007) due to the
tendency for the automatic visibility modelfitting to go awry for
complicated sources.

\begin{figure}
\plotone{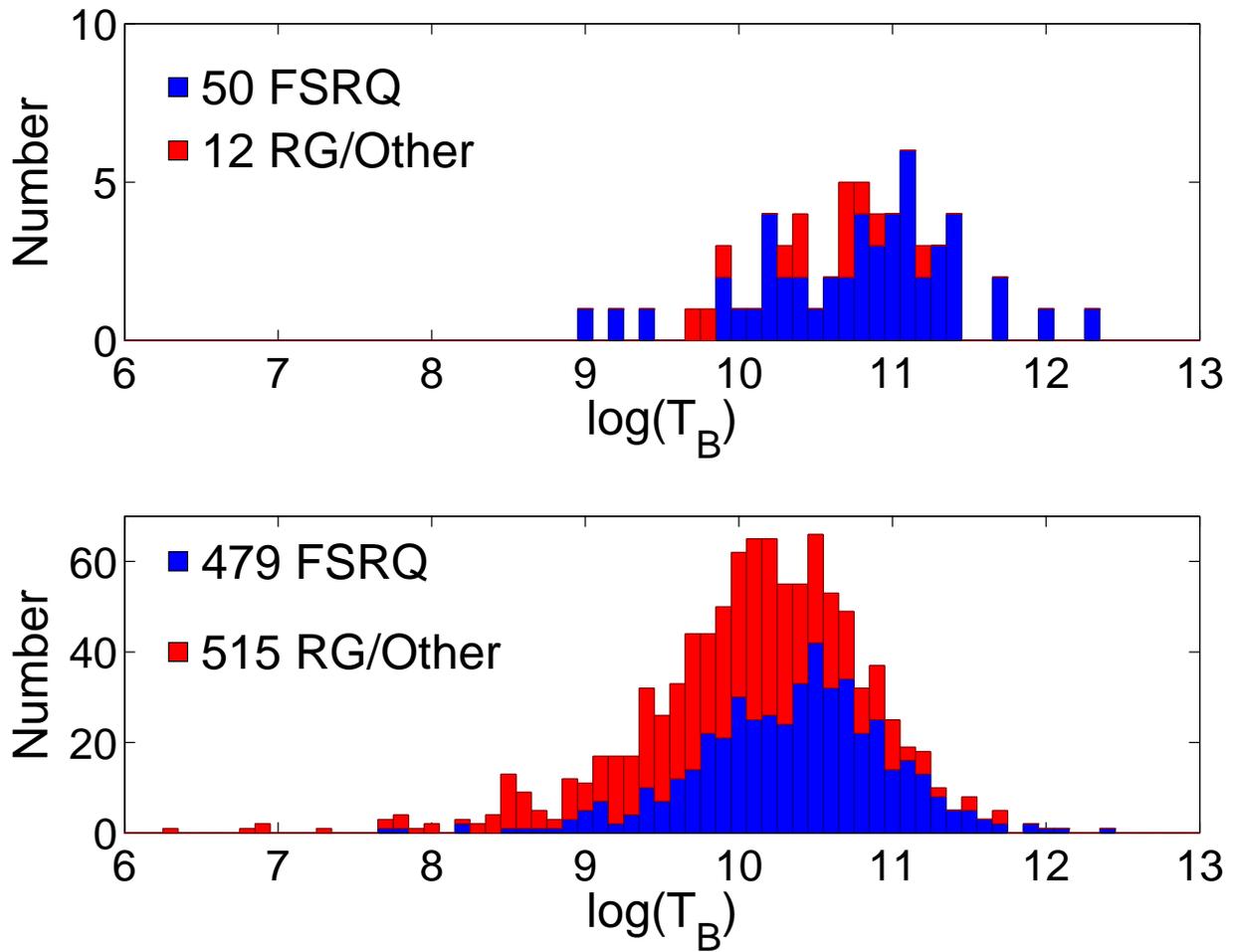}
\caption{
The distributions of core brightness temperatures for LAT-detected
FSRQs and radio galaxies/unclassified objects (top) compared to the distributions of core brightness temperature
for non-LAT FSRQs and radio galaxies/unclassified objects(bottom).}
\label{FSRQnOTb}
\end{figure}

There is strong evidence that the brightness temperature for the cores
of the LAT-detected sources may be higher than the $\gamma$-ray quiet population (see the FSRQ and radio galaxy/unclassified distributions in Fig.~\ref{FSRQnOTb}).
The BL Lac objects are the exception to this trend.  The median core brightness temperature for BL Lac objects detected by LAT is about $2.6 \times 10^{10}$ K, while the median for remaining non-LAT BL Lac objects is $2.7 \times 10^{10}$ K.  
The K-S test probability for these two distributions coming from the same
parent population is 93.6\%.  The median core brightness temperature for FSRQ objects in the LAT sample is $8.2 \times 10^{10}$ K, while the non-LAT FSRQ median is only about $2.5 \times 10^{10}$ K.  The K-S test probability that the two FSRQ distributions come from the same parent population is very low (0.001\%).  The median core brightness temperature for LAT radio galaxies and unclassified objects is $3.9 \times 10^{10}$ K, and for non-LAT objects the median is $1.0 \times 10^{10}$ K.  The K-S test probability the these distributions are drawn from the same parent sample is only 4.4\%.

Interestingly, $\gamma$-ray flux does not appear to have any relation to core brightness temperature.  
We applied the Spearman test to the $\gamma$-ray vs. core brightness temperature distributions and found correlation coefficients of 0.093 for BL Lacs, 0.111 for FSRQs, and -0.181 for radio galaxies and unclassified objects.

\subsection{Jet Opening Angle}

\begin{figure}
\plotone{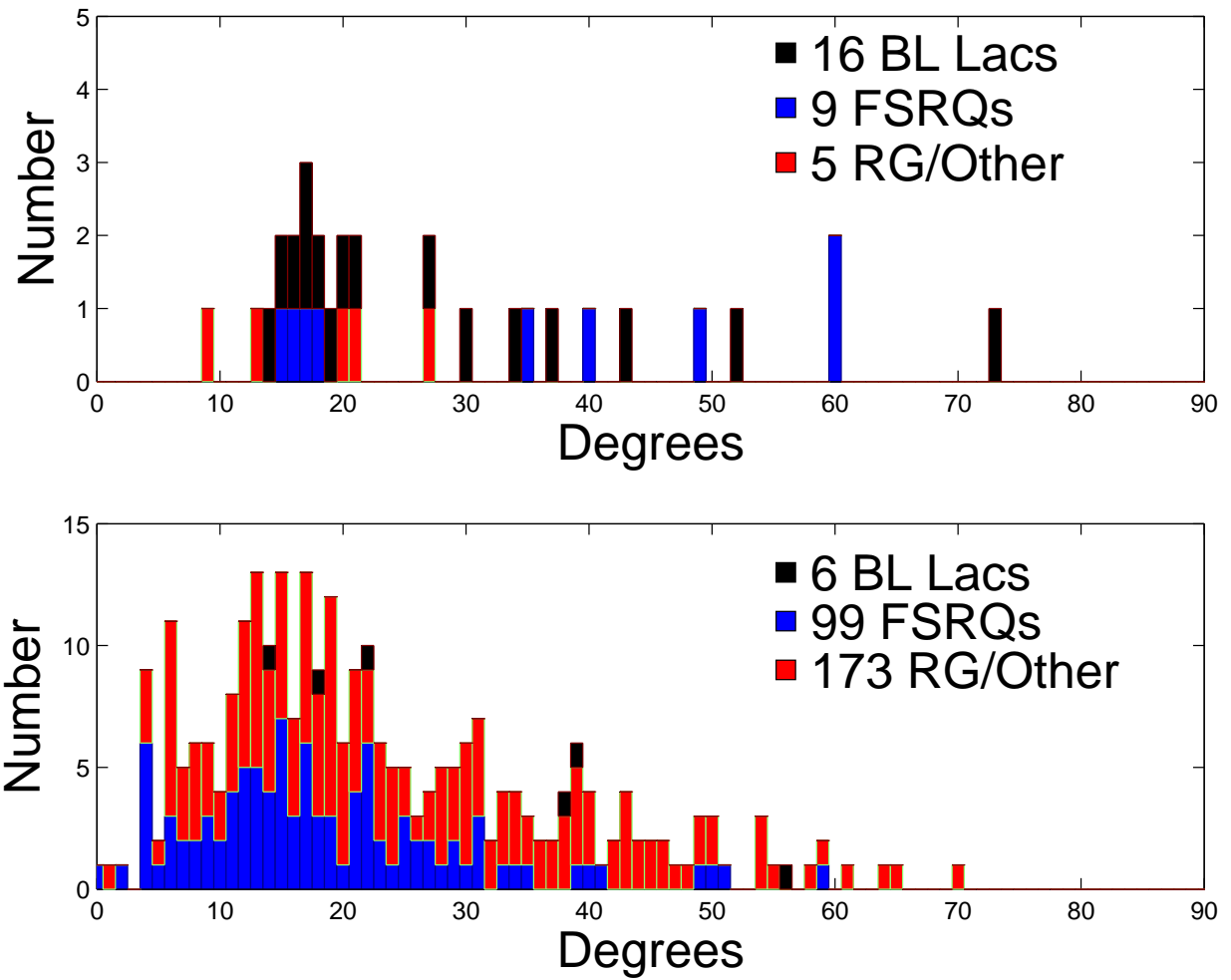}
\caption{
The distribution of opening angles for LAT-detected
BL Lacs, FSRQs and radio galaxies/unclassified objects (top) compared to the distribution of opening angles
for non-LAT objects (bottom).  
}
\label{angle_stack}
\end{figure}

We have measured a mean opening half-angle by the following 
procedure (Taylor et al.\ 2007):  We measure the separation of each jet component from the core along the
jet axis (taken to be a linear fit to the component positions) and the
distance of each component from the jet axis, i.e., x' and y'
positions in a rotated coordinate system with the jet axis along the
x'-axis.  For each component, we measure its extent from its center
along a line perpendicular to the jet axis using the parameters of its
elliptical fit, and then deconvolve this using the extent of the
Gaussian restoring beam along the same line.  The opening half-angle
measured from each component is then taken to be 
$$
\psi = {\rm arctan}[(|y'|+dr)/|x'|]
$$
 where $dr$ is the deconvolved Gaussian size
perpendicular to the jet axis.  After measuring this for each jet
component, we average them to get a single value.  This is only done
for sources with more than 2 total components, (i.e., at least 2 jet
components).  Unfortunately, we can only
make quantitative estimates for 30 of the LAT sources; 16 BL Lac, 9 FSRQ, and 5 radio galaxy or unclassified objects.
The distributions of opening angles are shown in Fig.~\ref{angle_stack}.  
Although this is based on very small
statistics there is evidence that the LAT source jets have unusually
large opening angles.  This is especially true for the FSRQ detected by LAT.  The MOJAVE (Pushkarev et al. 2009) and TANAMI (Ojha et al. 2010) groups both reported larger opening angles for $\gamma$-ray loud blazars.
Taylor et al.\ (2007) also found evidence for larger opening angles in EGRET sources.

We found 10 LAT-detected objects with opening angles greater than 30 degrees.  The large opening angle objects are BL Lacs and FSRQs, 5 of each.  See Fig.~\ref{loafigtile} for contour maps of these 10 sources.

\subsection{Core Fraction}

We define the core fraction as the ratio of flux density in the core compared to the
total flux density at 5 GHz.  A high core fraction can be an indication of a high Doppler factor.  Taylor et al.\ (2007) reported that EGRET sources tended to have greater core fractions than sources not detected by EGRET.  With the LAT sources, we find that this is not the case.  The median core fractions are 0.868 for LAT BL Lacs and 0.896 for non-LAT BL Lacs, 0.939 for LAT FSRQs and 0.929 for non-LAT FSRQs, and 0.889 for LAT radio galaxies/unclassified objects and 0.886 for non-LAT radio galaxies/unclassified objects.
The K-S test probabilities that the distributions were
drawn from the same parent distribution is 26.4\% for BL Lacs, 16.1\% for FSRQs and 23.5\% for radio galaxies/unclassified objects.

\subsection{Jet Bending, Length, and Brightness Temperature}

We can also compare the amount of bending in the jet on the 
parsec scale for the LAT sources compared to non-LAT sources.  Unfortunately, as with the jet opening angle measurements, we are limited to only 30 LAT sources  
and the results are inconclusive.  

\begin{figure}
\plotone{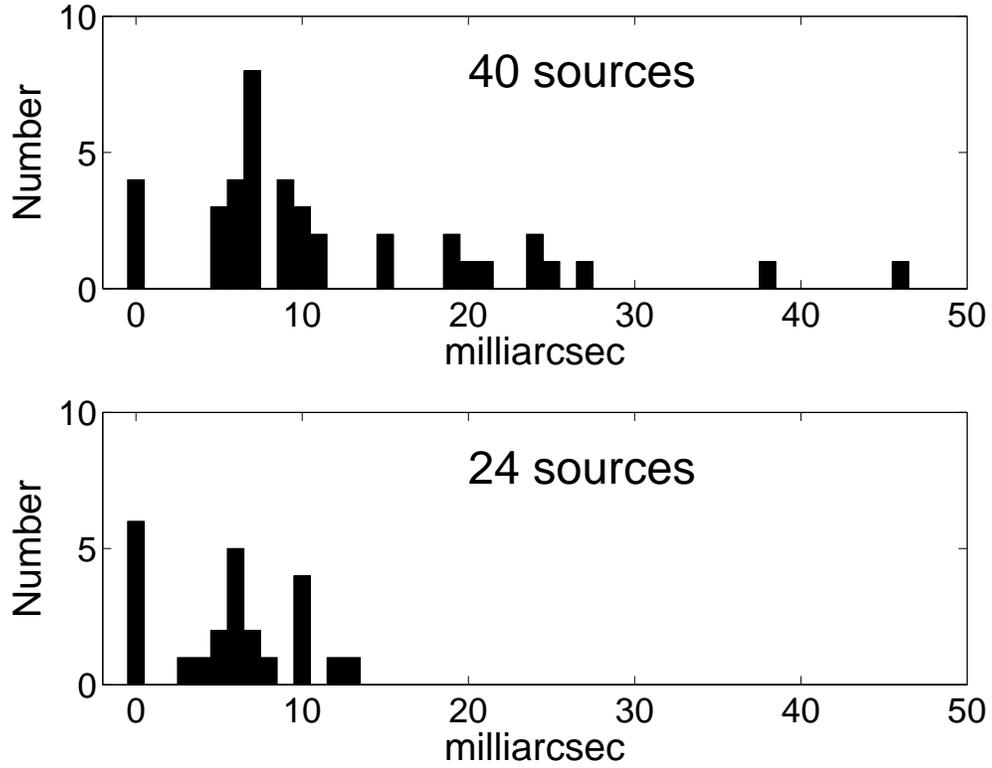}
\caption{
The distribution of jet lengths for LAT-detected
BL Lac (top) compared to the distribution of jet lengths
for non-LAT BL Lac objects (bottom).  
}
\label{jetlen_bll}
\end{figure}

We compared
the distribution of jet length in LAT sources with non-LAT sources. 
There does not appear to be an appreciable difference between the two
populations of FSRQ an radio galaxies/unclassified objects.  The K-S test results are 63.3\% for the FSRQ objects, and 64.6\% for the radio galaxy and unclassified objects.  Interestingly, the BL Lac objects in the LAT sample appear to have longer jets than the non-LAT BL Lacs.  The distributions for the BL Lacs are shown in Fig.~\ref{jetlen_bll}.  The K-S test probability that the two are drawn from the same distribution is 2.1\%.  The FSRQ and AGN/other objects have essentially the same distributions.  

Unlike the core brightness temperatures, the jet brightness temperatures
(formally, the brightness temperature of the brightest jet component) for the LAT and non-LAT distributions
look fairly similar. The K-S test probabilities for the BL Lac, FSRQ, and radio galaxy/other pairs of distributions are 34.2\%, 92.1\%, and 83.5\% respectively.  

\section{Notes on Individual Sources}

\subsection{Low Core Brightness Temperature}

\begin{figure}
\epsscale{0.8}
\plotone{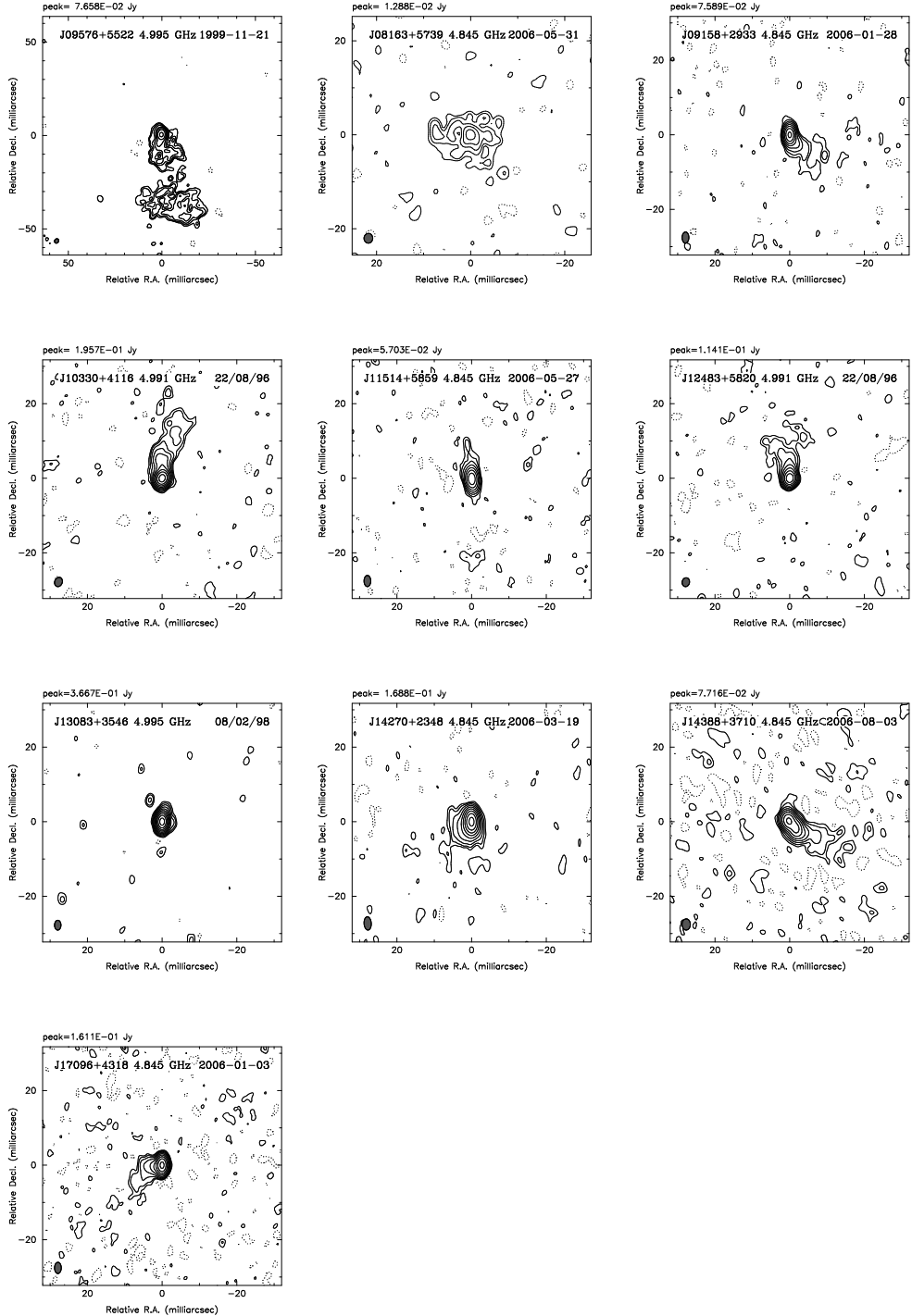}
\caption{
Contour maps of the 10 LAT sources with opening angles larger than $30^\circ$.  All data was taken with the VLBA at 5 GHz.  Bottom contours are at 0.5 mJy/beam, except for the first object which has a bottom contour level is 1 mJy/beam.  The restoring beam is shown as an ellipse in the lower left corner of each panel.
}
\label{loafigtile}
\end{figure}

\noindent
{\bf J09576+5522} With a core brightness temperature of only $9.7 \times 10^{8}$ K, this FSRQ has a $\gamma$-ray flux of $106.5 \pm 7.4 \times 10^{-9}$ photons cm$^{-2}$ s$^{-1}$, nearly three times the median for FSRQs.  Also known as [HB89] 0954+556, it has a redshift of 0.896 (Abazajian et al.\ 2004) and a 8.5 GHz flux density is 1.5 Jy.  This FSRQ has been studied in the UV (Pian, Falomo, \& Treves 2005), radio, X-ray (Tavecchio, et al.\ 2007), and was detected in the $\gamma$-rays by EGRET (Hartman et al.\ 1999).  Tavecchio et al.\ (2007) noted that X-ray flux was only found at the terminal portion of the jet.  The automated program initially reported the core brightness temperature for this object to be $1.2 \times 10^{8}$ K.  We suspected program did not generate an accurate model for this source.  When we did the model fitting by hand in DIFMAP, we calculated a core brightness temperature of $9.7 \times 10^{8}$ K.  This is the lowest core temperature of all objects in our LAT sample.  This object is featured in Fig.~\ref{loafigtile} because its jet opening angle is about $60^\circ$.  

\noindent
{\bf J08163+5739} The automated program found that this BL Lac had a core brightness temperature of only $4.16 \times 10^{7}$ K.  Again, when we did the model-fitting by hand we calculated a core brightness temperature of $1.6 \times 10^{9}$ K.  Even with this higher core brightness temperature, this object is still the coolest BL Lac in our LAT sample.  This object has a $\gamma$-ray flux of $15.2 \pm 0.3 \times 10^{-9}$ photons cm$^{-2}$ s$^{-1}$.   
It is a compact and weak radio source with a radio flux density of 92 mJy at 8.5 GHz.  There is no measured redshift for this object.  It is also known as SBS 0812+578 and 1RXS J081624.6+573910.  It has a small jet extending to the east.   
This object is featured in Fig.~\ref{loafigtile} because its jet opening angle is approximately $52^\circ$.

\subsection{High Redshift}

\noindent
{\bf J07464+2549} With a redshift of 2.979 (Abazajian et al.\ 2004), this FSRQ has a $\gamma$-ray flux $42.3 \pm 7.9 \times 10^{-9}$ photons cm$^{-2}$ s$^{-1}$.  Its core brightness temperature is $5.07 \times 10^{11}$ K, which is well above the median for the LAT-detected FSRQs.  It is also known as B2 0743+25 and RX J0746.4+2549.  It is a relatively compact and bright radio source with a flux density of 731 mJy at 8.5 GHz (see Fig.~\ref{indisour}).  It has a very small jet extending to the north.  In 2005, the Burst Alert Telescope on board the \textit{Swift} satellite detected a flare of 15-195 keV emission from this object (Sambruna et al.\ 2006).

\noindent
{\bf J08053+6144} With a redshift of 3.033 (Sowards-Emmerd et al.\ 2005), this FSRQ has the highest redshift among the LAT-detected objects.  It has a $\gamma$-ray flux $36.0 \pm 8.9 \times 10^{-9}$ photons cm$^{-2}$ s$^{-1}$.  Its core brightness temperature is $1.30 \times 10^{11}$ K, which is slightly higher than the median for LAT-detected FSRQs.  
This object is fairly bright in the radio with a flux density of 725 mJy at 8.5 GHz.  It has a short jet extending to the southeast (see Fig.~\ref{indisour}).

\subsection{Radio Galaxies}

Our sample contains six sources identified as radio galaxies, although in some cases there is controversy in this identification as described below.

\begin{figure}
\epsscale{0.9}
\plotone{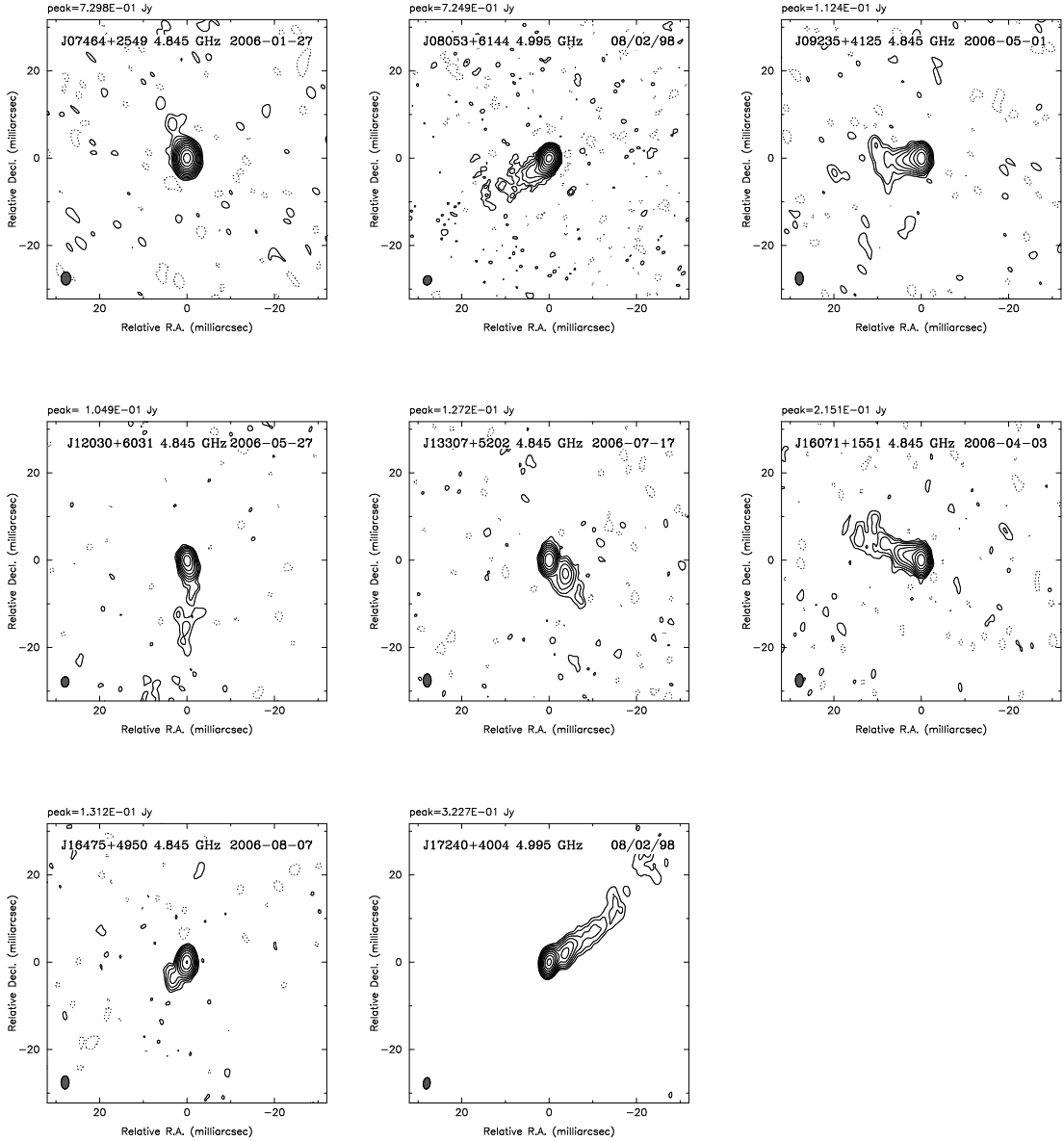}
\caption{
Contour maps of the objects in the sections 5.2 and 5.3.  All data was taken with the VLBA at 5 GHz.  Bottom contours are at 0.5 mJy/beam.  The restoring beam is shown as an ellipse in the lower left corner of each panel.
}
\label{indisour}
\end{figure}

\noindent
{\bf J09235+4125} This object is also known as B3 0920+416 and GB6 J0923+4125.  It has a $\gamma$-ray flux of $40.6 \pm 10.0 \times 10^{-9}$ photons cm$^{-2}$ s$^{-1}$ and a redshift of 0.028 (Falco, Kochanek, \& Mu\~noz 1998).  Its peak radio flux density is 120 mJy at 8.5 GHz.  It has a short jet extending to the east (see Fig.~\ref{indisour}).  The NASA/IPAC Extragalactic Database (NED) classifies this object as a FSRQ.  However, this is probably an error.  Falco, Kochanek, \& Mu\~noz (1998) call it a ``late-type galaxy.''

\noindent
{\bf J12030+6031} This object is also known as SBS 1200+608 and RX J1203.0+6031.  It has a $\gamma$-ray flux of $44.8 \pm 0.3 \times 10^{-9}$ photons cm$^{-2}$ s$^{-1}$ and a redshift of 0.065 (Falco, Kochanek, \& Mu\~noz 1998).  Its peak radio flux density is 105 mJy at 8.5 GHz.  It has a short jet extending to the south (see Fig.~\ref{indisour}).  V\'eron-Cetty \& V\'eron (2006) classify this object as a low ionization nuclear emission region (LINER) galaxy, sometimes referred to as a Seyfert 3.

\noindent
{\bf J13307+5202} This radio galaxy is also known as CGRaBS J1330+5202.  It has an upper limit to its $\gamma$-ray flux of $41.5 \times 10^{-9}$ photons cm$^{-2}$ s$^{-1}$ and a redshift of 0.688 (Healey et al.\ 2008).  It has a peak radio flux density of 127 mJy at 8.5 GHz and a jet extending to the southwest (see Fig.~\ref{indisour}).  Sowards-Emmerd et al.\ (2005) list it as an unknown optical type.

\noindent
{\bf J16071+1551} This object is also known as [HB89] 1604+159.  It is an EGRET-detected source known as 3EG J1605+1553.  It has a LAT $\gamma$-ray flux of $33.9 \pm 7.7  \times 10^{-9}$ photons cm$^{-2}$ s$^{-1}$ and a redshift of 0.496 (Adelman-McCarthy et al. 2008).  It has a peak radio flux density of 214 mJy at 8.5 GHz and a jet extending to the east (see Fig.~\ref{indisour}).  V\'eron-Cetty \& V\'eron (2006) classify this object as a BL Lac.

\noindent
{\bf J16475+4950} This object is also known as SBS 1646+499 and 1RXS J164735.4+495001.  It has a $\gamma$-ray flux of $33.3 \pm 8.5 \times 10^{-9}$ photons cm$^{-2}$ s$^{-1}$ and a redshift of 0.047 (Falco, Kochanek, \& Munoz 1998).  It has a peak radio flux density of 131 mJy at 8.5 GHz.  It has a very short jet extending to the southeast (see Fig.~\ref{indisour}).  The spectrum in Marcha et al.\ (1996) indicates that this object is a Seyfert 1 blazar.

\noindent
{\bf J17240+4004} This object is also known as B2 1722+40 and RX J1724.0+4004.  It has a $\gamma$-ray flux of $47.1 \pm 9.0 \times 10^{-9}$ photons cm$^{-2}$ s$^{-1}$ and a redshift of 1.049 (Vermeulen, Taylor, \& Readhead 1996).  This is the brightest radio galaxy in our sample with a peak flux density of 323 mJy at 8.5 GHz.  It has a long jet extending to the northwest (see Fig.~\ref{indisour}).  V\'eron-Cetty \& V\'eron (2006) tentatively classify this object as a BL Lac.  However, the spectrum in Vermeulen, Taylor, \& Readhead (1996) does not support this classification.  The spectrum has only narrow lines, so it cannot be a FSRQ, and the equivalent widths are greater than 5 angstroms, so it cannot be a BL Lac.
 
\section{Discussion}

One of the primary explanations for some AGN being $\gamma$-ray loud is Doppler Boosting.  The orientation and material speed of the AGN jets have a large effect on their observed fluxes due to the Doppler effect.  The kinematic Doppler factor for a moving source is defined as 
$$
\delta = [\Gamma(1 - \beta \cos \theta)]^{-1}
$$
where $\beta$ is the bulk velocity $v/c$, $\Gamma$ is the Lorentz factor, $\Gamma=(1-\beta^{2})^{-1/2}$, and $\theta$ is the angle between the velocity vector and the line of sight.  Emissions from an approaching source are blue-shifted by
$$
\nu = \delta\nu'
$$
where $\nu'$ is the frequency in the source's rest frame.  The intensity enhancement, commonly known as the ``Doppler boosting'', is given in Rybicki and Lightman (1979) as
$$
I_{\nu}(\nu) = \delta^{3}I'_{\nu'}(\nu').
$$
Because LAT measures the flux across a large bandwidth (1 to 100 GeV), we have to integrate over frequency:
$$
F = \int I_{\nu} d\nu = \int \delta^{3}I'_{\nu'} \delta d\nu' = \delta^{4}F'
$$ 
Therefore, a small difference in material velocity or orientation angle leads to a large difference in intensity.  See Urry \& Padovani (1995) for a more thorough discussion of Doppler enhancement.  

\subsection{BL Lacs}

Almost two-thirds of the BL Lac objects in VIPS are detected by LAT, and the only significant difference between the LAT and non-LAT populations is in their jet lengths. We have shown that the LAT BL Lacs have longer jets than the non-LAT BL Lacs (see \S4.6).  
There are no differences in the fraction of polarized BL Lacs or the distributions of the polarization.  The LAT and non-LAT BL Lacs have nearly identical core brightness temperature distributions.  Why are the non-LAT BL Lacs quiet in the $\gamma$-ray?  It is possible that the non-LAT BL Lacs are fundamentally different from the LAT BL Lacs, but it must be in some subtle way.  It seems more likely that all the BL Lacs are emitting in the $\gamma$-ray, but some are emitting below the threshold for the LAT to detect.

Initially, we suspected that the non-LAT BL Lacs might have been too far away to detect their $\gamma$-rays.  
The more distant BL Lacs definitely have lower $\gamma$-ray flux, but it is clear that the non-LAT BL Lacs are not at higher redshifts than the LAT BL Lacs.  While we only have redshift measurements for 16 of the LAT BL Lacs and 15 of the non-LAT BL Lacs, that is still enough to rule out redshift as the cause of LAT non-detection.  The maximum measured redshift for the LAT BL Lacs is 0.56.  Nearly half ($11/24$) of the non-LAT BL Lacs have redshifts lower than 0.56.  Also, there is no strong correlation between redshift and $\gamma$-ray flux for the BL Lacs (see \S3.1).  

The prevalence of long-jet BL Lacs is an interesting clue to this puzzle.  The K-S test result of 2.1\% for the jet length distributions (see \S4.6) is only of modest statstical significance, but the fact that a larger percentage of the LAT BL Lacs are classified as LJET than the non-LAT BL Lacs (see Table~\ref{morphtable}) supports it.  One would expect the jets to appear shorter if they are pointed more directly at us.  However, if the $\gamma$-ray loud BL Lacs have higher Doppler Factors than their $\gamma$-ray quiet counterparts, that would indicate that the LAT BL Lacs have more powerful jets than the non-LAT BL Lacs.  A more powerful jet will appear longer even when viewed at a small angle.

Because we find that the LAT BL Lacs have longer jets than non-LAT BL Lacs, the orientation angle for LAT BL Lacs cannot be smaller than non-LAT BL Lacs.   The non-LAT BL Lacs could have slightly lower jet speeds.  This can explain why the LAT BL Lacs have longer jets and why they are $\gamma$-ray loud.  Unfortunately, this theory is not supported by recent work by Lister et al.\ (2009b), where they find that two non-LAT BL Lacs have higher maximum jet material velocities than LAT BL Lacs.  However, they only had 10 BL Lacs in their LAT-detected sample, so this is still an open question.

BL Lacs are known to have high variability in their flux densities.  The non-LAT BL Lacs could currently be in low $\gamma$-ray luminosity states, but may enter states of enhanced $\gamma$-ray activity at a later time.  It is possible that the jet material speed is related to this flux variability.  Based on the number of BL Lacs in our sample, the duty cycle would need to be about $2/3$ over the course of 1 year.

\subsection{FSRQs}

The LAT FSRQs seem to be substantially different from the non-LAT FSRQs.  Only about 9\% (50/529) of the FSRQs in VIPS are $\gamma$-ray loud.  This indicates the LAT FSRQs are special in some way.  Recall from \S3.2 that $\gamma$-ray flux correlates with radio flux density for FSRQs, especially for high flux density sources.  Also, in general the LAT FSRQs have significantly higher core brightness temperatures, with the notable exception of J09576+5522, and they appear to have larger opening angles.  So, brighter FSRQs are more likely to produce $\gamma$-rays.  We showed in \S3.1 that $\gamma$-ray flux does not correlate with redshift, so we do not expect the LAT FSRQs to have lower redshift than their non-LAT counterparts.  We explored the possibility that LAT FSRQs were at low redshift, but found that they cover a range similar to the non-LAT FSRQs.    

The $\gamma$-ray flux from FSRQs can be explained with Doppler boosting, but requires a more dramatic difference than with the BL Lacs.  
Lister et al.\ (2009b) reported that LAT-detected FSRQs had higher median jet speeds than LAT-detected BL Lacs by more than a factor of two.  It is also possible that there are two subclasses for FSRQs: $\gamma$-ray loud and $\gamma$-ray quiet.  The difference in $\gamma$-ray detection rates for FSRQs could also be due to variations in the $\gamma$-ray emission, however the duty cycle required (about 10\%) is much more extreme than for the BL Lacs.

A larger fraction of the LAT FSRQs have polarization in their cores compared to the non-LAT FSRQs.  We detected polarization in the cores of 48\% of the LAT FSRQs compared to 33\% of the non-LAT FSRQs, a difference that is significant at about the 1-sigma level.  
This points to an intrinsic difference in $\gamma$-ray quiet versus $\gamma$-ray loud FSRQs.  The higher polarization could indicate more organized magnetic fields, possibly pointing to a greater spin of the supermassive black hole (Meier 2001).  A higher black hole spin would also lead to higher material velocities in the jets (Meier 1999).

\subsection{Radio Galaxies}

There is a definite lack of CSOs among the LAT-detected objects.  This is not surprising from a Doppler boosting point of view.  For an object to appear symmetric, we need to see it close to ``edge on'' so we can see both jets.  With the jets pointed nearly perpendicular to our line of sight, we would not expect to see much Doppler boosting in the radiation.  However, Stawarz et al.\ (2008) suggested that CSOs should account for a ``relatively numerous class of extragalactic sources'' for the \textit{Fermi Gamma-ray Space Telescope} due to ultrarelativistic electrons in the lobes of these compact objects.  We have yet to detect any $\gamma$-rays from a CSO.

The LAT radio galaxies and unclassified objects are a puzzle.  They do not appear to differ significantly from their non-LAT counterparts in any appreciable way.  Their opening angles appear to be very similar to the non-LAT objects and they are not particularly bright in the radio.  The LAT radio galaxies/unclassified objects may have higher core temperatures than the non-LAT objects.  The one difference that stands out is that we detected polarization in the cores of 67\% of the LAT radio galaxies/unclassified objects compared to only 20\% for their non-LAT counterparts.  Unfortunately, our current sample is too small to make any definite claims about these objects.  There is also controversy about the optical types of several of the objects we label as radio galaxies.

\subsection{Constraints on the $\gamma$-ray Emitting Region}

The major differences between LAT and non-LAT sources are confined to the cores of the sources.  The LAT BL Lacs have stronger jets, suggesting higher ejection velocities from the cores.  The LAT FSRQs have higher core temperatures and more frequently exhibit polarized cores.  The LAT radio galaxies/unclassified objects are also more polarized than their non-LAT counterparts.  We suspect that the $\gamma$-rays are being produced very near to the cores of these objects.  To put a limit on the size of the $\gamma$-ray emitting region, we measured the sizes of the cores for several objects.  Using all of the objects for which we have measured redshifts, the median upper bound on the core size is 4.54 pc with a standard deviation of 10.98 pc.  However, a better estimate for the upper limit on the size of the emitting region should come from measurements of the objects nearest to us.  Limiting ourselves to objects with $z\leq0.1$, we have 4 BL Lacs and 3 radio galaxies (a LINER, a Seyfert 1, and an object that may actually be a FSRQ).  Using this subset of sources, we find a median upper bound on the core size of 0.58 pc with a standard deviation of 0.34 pc.   
Therefore, the $\gamma$-ray emission region must be within about 0.9 parsecs of the central engine.

\section{Conclusions}

We compared the distributions of core polarization, core brightness temperature, jet opening angles, core fractions, jet bending, jet length, and 8.5 GHz radio flux densities for LAT and non-LAT sources. The only significant difference between the LAT and non-LAT BL Lacs is that the $\gamma$-ray loud BL Lacs appear to have longer jets.  Also, we find no correlation between BL Lac radio flux density and $\gamma$-ray flux.  We suspect that all BL Lacs are producing $\gamma$-rays, but some are below LAT's threshold due to jet material speed, or possibly variability in the $\gamma$-ray luminosity.  In contrast, $\gamma$-ray bright FSRQs are considerably different from the $\gamma$-ray quiet population.  The LAT FSRQs appear to be extreme sources with higher core brightness temperatures, larger opening angles, and greater core polarization than their non-LAT counterparts.  There is also a significant correlation between the FSRQ radio flux density and $\gamma$-ray flux, especially for brighter sources.  The $\gamma$-ray loud FSRQs can be explained by Doppler boosting, but the orientation and/or jet material speeds must be significantly different than the $\gamma$-ray quiet FSRQs.  One possible way to explain this difference is that the $\gamma$-ray loud FSRQs may have greater angular momentum in their central supermassive black holes, leading to higher jet velocities and greater polarization.  
Due to the small number of radio galaxies in our sample, combined with the controversy over some of their optical types, there is little we can say about $\gamma$-ray bright radio galaxies.  However, we do find two intriguing clues about them.  First, there are no $\gamma$-ray loud CSOs in our sample.  Second, the radio galaxies that we do detect in the $\gamma$-rays tend to have strong polarizations in their cores.

Because the differences we find between the $\gamma$-ray loud and $\gamma$-ray quiet objects are related to the core, we suspect that the $\gamma$-ray radiation originates within the core (i.e., at the base of the jet).  By calculating the core sizes of the LAT objects with measured redshifts, we put an upper limit of 15 pc on the size of this emitting region.  A better limit of 0.9 pc results from using the core sizes of the lowest redshift objects in our LAT sample.

As more data is collected more will become clear about $\gamma$-ray loud blazars.  We, and several other groups, are currently conducting further observations.  We will soon have more information about the velocities of the jet material and more associations of AGN with $\gamma$-ray sources.

\noindent \textbf{Acknowledgements}

\noindent We would like to thank Matt Lister and Ann Wehrle for useful discussions.  We would also like to thank Teddy Cheung, Marcello Giroletti, Yuri Kovalev, Frank Schinzel, David Thompson, and Steve Tremblay for their helpful comments.  Also, we would like to express our appreciation to the anonymous referee for their insightful and helpful criticism.  The National Radio Astronomy Observatory is a facility of the National Science Foundation operated under cooperative agreement by Associated Universities, Inc.  The NASA/IPAC Extragalactic Database (NED) is operated by the Jet Propulsion Laboratory, California Institute of Technology, under contract with the National Aeronautics and Space Administration.  We thank NASA for support under FERMI grant GSFC \#21078/FERMI08-0051.


\textwidth = 7.0truein
\textheight = 10.0truein
\begin{deluxetable}{ccrrrcrrrrrrr}
\tablecolumns{13}
\tabletypesize{\scriptsize}
\tablewidth{0pt}
\rotate
\tablecaption{LAT Sources in VIPS}
\tablehead{
\colhead{1FGL Name}	&	\colhead{VIPS Name}	&	\colhead{RA}	&	\colhead{DEC}	&	\colhead{Prob}	&	\colhead{Type}	&	\colhead{z}	&	\colhead{F$_{35}$}	&	\colhead{$\Delta$ F$_{35}$} & \colhead{Core T$_B$} & \colhead{Open. Ang.} & \colhead{$\Delta$ PA} & \colhead{S (8.5 GHz)} \\}
\startdata
1FGL J0742.2+5443	&	J07426+5444	&	07:42:39.8	&	+54:44:24.679	&	1	&	FSRQ	&	0.723	&	2.2	&	0.3	&	5.26E+10	&	\nodata	&	\nodata	&	142.8	\\
1FGL J0746.6+2548	&	J07464+2549	&	07:46:25.9	&	+25:49:02.146	&	0.98	&	FSRQ	&	2.979	&	0.7	&	0.2	&	5.07E+11	&	\nodata	&	\nodata	&	291.7	\\
1FGL J0752.8+5353	&	J07530+5352	&	07:53:01.4	&	+53:52:59.636	&	1	&	BLL	&	0.2	&	1.2	&	0.3	&	6.79E+11	&	\nodata	&	\nodata	&	1196.9	\\
1FGL J0800.5+4407	&	J08011+4401	&	08:01:08.3	&	+44:01:10.148	&	0.92	&	\nodata	&	\nodata	&	0.7	&	0.3	&	7.36E+10	&	\nodata	&	\nodata	&	230.6	\\
1FGL J0806.2+6148	&	J08053+6144	&	08:05:18.2	&	+61:44:23.704	&	0.94	&	FSRQ	&	3.033	&	0.6	&	0.2	&	1.30E+11	&	\nodata	&	\nodata	&	725.7	\\
1FGL J0809.4+3455	&	J08096+3455	&	08:09:38.9	&	+34:55:37.248	&	0.99	&	BLL	&	0.082	&	0.7	&	0.3	&	2.48E+10	&	\nodata	&	\nodata	&	152.1	\\
1FGL J0809.5+5219	&	J08098+5218	&	08:09:49.2	&	+52:18:58.252	&	0.99	&	BLL	&	0.138	&	2.4	&	0.3	&	4.83E+09	&	\nodata	&	\nodata	&	154.2	\\
1FGL J0815.0+6434	&	J08146+6431	&	08:14:39.2	&	+64:31:22.040	&	0.99	&	BLL	&	\nodata	&	1.7	&	0.3	&	2.79E+10	&	\nodata	&	\nodata	&	221.6	\\
1FGL J0816.7+5739	&	J08163+5739	&	08:16:23.8	&	+57:39:09.509	&	1	&	BLL	&	\nodata	&	0.7	&	0.3	&	4.16E+07	&	52	&	8	&	92	\\
1FGL J0818.2+4222	&	J08182+4222	&	08:18:16.0	&	+42:22:45.408	&	1	&	BLL	&	\nodata	&	8.7	&	0.6	&	3.76E+10	&	\nodata	&	\nodata	&	1046.4	\\
1FGL J0825.0+5555	&	J08247+5552	&	08:24:47.2	&	+55:52:42.662	&	0.99	&	FSRQ	&	1.417	&	0.9	&	0.3	&	7.86E+10	&	15	&	-7	&	1667.8	\\
1FGL J0834.4+4221	&	J08338+4224	&	08:33:53.9	&	+42:24:01.859	&	0.98	&	FSRQ	&	0.249	&	0.9	&	0.2	&	8.66E+09	&	\nodata	&	\nodata	&	560.8	\\
1FGL J0856.6+2103	&	J08566+2057	&	08:56:39.7	&	+20:57:43.426	&	0.94	&	BLL	&	\nodata	&	1	&	0.3	&	8.98E+09	&	\nodata	&	\nodata	&	132.7	\\
1FGL J0856.6+2103	&	J08569+2111	&	08:56:57.2	&	+21:11:43.640	&	0.96	&	FSRQ	&	2.098	&	1	&	0.3	&	1.70E+10	&	\nodata	&	\nodata	&	332.8	\\
1FGL J0910.7+3332	&	J09106+3329	&	09:10:37.0	&	+33:29:24.418	&	1	&	BLL	&	0.354	&	0.8	&	0.3	&	7.40E+10	&	\nodata	&	\nodata	&	88.6	\\
1FGL J0911.0+2247	&	J09107+2248	&	09:10:42.1	&	+22:48:35.565	&	0.98	&	FSRQ	&	2.661	&	2	&	0.3	&	8.82E+10	&	\nodata	&	\nodata	&	178.8	\\
1FGL J0912.3+4127	&	J09121+4126	&	09:12:11.6	&	+41:26:09.356	&	0.98	&	FSRQ	&	2.563	&	0.6	&	0.2	&	8.06E+09	&	\nodata	&	\nodata	&	166.9	\\
1FGL J0915.7+2931	&	J09158+2933	&	09:15:52.4	&	+29:33:23.982	&	1	&	BLL	&	\nodata	&	2.1	&	0.3	&	1.87E+10	&	37	&	6	&	186.1	\\
1FGL J0919.6+6216	&	J09216+6215	&	09:21:36.2	&	+62:15:52.185	&	0.82	&	FSRQ	&	1.446	&	1.1	&	0.3	&	2.61E+11	&	\nodata	&	\nodata	&	1532	\\
1FGL J0920.9+4441	&	J09209+4441	&	09:20:58.5	&	+44:41:53.988	&	1	&	FSRQ	&	2.19	&	14	&	0.7	&	8.58E+10	&	\nodata	&	\nodata	&	1368.3	\\
1FGL J0923.2+4121	&	J09235+4125	&	09:23:31.3	&	+41:25:27.429	&	0.97	&	RG	&	0.028	&	1.2	&	0.3	&	4.91E+09	&	27	&	-8	&	235.2	\\
1FGL J0924.2+2812	&	J09238+2815	&	09:23:51.5	&	+28:15:24.966	&	0.97	&	FSRQ	&	0.744	&	1.3	&	0.3	&	9.87E+11	&	\nodata	&	\nodata	&	218.4	\\
1FGL J0929.4+5000	&	J09292+5013	&	09:29:15.4	&	+50:13:35.982	&	0.67	&	BLL	&	\nodata	&	0.7	&	0.2	&	6.67E+10	&	\nodata	&	\nodata	&	747.8	\\
1FGL J0934.5+3929	&	J09341+3926	&	09:34:06.7	&	+39:26:32.125	&	0.93	&	BLL	&	\nodata	&	0.6	&	0.2	&	1.51E+10	&	\nodata	&	\nodata	&	188.3	\\
1FGL J0937.7+5005	&	J09372+5008	&	09:37:12.3	&	+50:08:52.082	&	0.97	&	FSRQ	&	0.276	&	0.8	&	0.2	&	6.26E+10	&	\nodata	&	\nodata	&	372.5	\\
1FGL J0941.2+2722	&	J09418+2728	&	09:41:48.1	&	+27:28:38.818	&	0.97	&	FSRQ	&	1.306	&	0.9	&	0	&	1.94E+10	&	\nodata	&	\nodata	&	268.4	\\
1FGL J0949.8+1757	&	J09496+1752	&	09:49:39.8	&	+17:52:49.432	&	0.98	&	FSRQ	&	0.693	&	0.9	&	0	&	1.27E+11	&	\nodata	&	\nodata	&	153	\\
1FGL J0956.9+2513	&	J09568+2515	&	09:56:49.9	&	+25:15:16.047	&	0.99	&	FSRQ	&	0.712	&	0.7	&	0.2	&	1.55E+10	&	\nodata	&	\nodata	&	1905	\\
1FGL J0957.7+5523	&	J09576+5522	&	09:57:38.2	&	+55:22:57.740	&	1	&	FSRQ	&	0.896	&	10.5	&	0.6	&	1.20E+08	&	60	&	-33	&	1498.9	\\
1FGL J1000.9+2915	&	J10011+2911	&	10:01:10.2	&	+29:11:37.543	&	0.99	&	BLL	&	\nodata	&	1.4	&	0.3	&	3.59E+10	&	\nodata	&	\nodata	&	262.1	\\
1FGL J1015.1+4927	&	J10150+4926	&	10:15:04.1	&	+49:26:00.704	&	1	&	BLL	&	0.2	&	8.7	&	0.6	&	8.16E+09	&	16	&	-1	&	252.2	\\
1FGL J1023.6+3937	&	J10231+3948	&	10:23:11.6	&	+39:48:15.378	&	0.89	&	FSRQ	&	1.254	&	0.6	&	0.2	&	9.53E+10	&	16	&	9	&	779.6	\\
1FGL J1033.2+4116	&	J10330+4116	&	10:33:03.7	&	+41:16:06.234	&	0.99	&	FSRQ	&	1.117	&	1.1	&	0.3	&	4.84E+10	&	49	&	163	&	381.4	\\
1FGL J1033.8+6048	&	J10338+6051	&	10:33:51.4	&	+60:51:07.342	&	1	&	FSRQ	&	1.401	&	2.2	&	0.3	&	1.16E+10	&	\nodata	&	\nodata	&	430.4	\\
1FGL J1043.1+2404	&	J10431+2408	&	10:43:09.0	&	+24:08:35.430	&	0.99	&	BLL	&	0.56	&	1.5	&	0.3	&	1.56E+10	&	\nodata	&	\nodata	&	685.4	\\
1FGL J1058.6+5628	&	J10586+5628	&	10:58:37.7	&	+56:28:11.180	&	1	&	BLL	&	0.143	&	5.7	&	0.5	&	9.49E+09	&	\nodata	&	\nodata	&	189.8	\\
1FGL J1104.4+3812	&	J11044+3812	&	11:04:27.3	&	+38:12:31.794	&	1	&	BLL	&	0.03	&	26.1	&	1	&	2.13E+11	&	15	&	-1	&	631.6	\\
1FGL J1106.5+2809	&	J11061+2812	&	11:06:07.3	&	+28:12:47.045	&	0.97	&	\nodata	&	\nodata	&	0.9	&	0.2	&	5.08E+10	&	\nodata	&	\nodata	&	193.2	\\
1FGL J1112.8+3444	&	J11126+3446	&	11:12:38.8	&	+34:46:39.124	&	0.99	&	FSRQ	&	1.949	&	0.9	&	0.3	&	6.95E+10	&	\nodata	&	\nodata	&	160.5	\\
1FGL J1123.9+2339	&	J11240+2336	&	11:24:02.7	&	+23:36:45.876	&	1	&	BLL	&	\nodata	&	0.9	&	0	&	1.07E+10	&	\nodata	&	\nodata	&	472.7	\\
1FGL J1141.8+1549	&	J11421+1547	&	11:42:07.7	&	+15:47:54.202	&	0.99	&	\nodata	&	\nodata	&	1	&	0.3	&	5.62E+10	&	20	&	-15	&	274	\\
1FGL J1146.8+4004	&	J11469+3958	&	11:46:58.3	&	+39:58:34.307	&	0.98	&	FSRQ	&	1.089	&	1	&	0.3	&	4.80E+11	&	\nodata	&	\nodata	&	571.5	\\
1FGL J1150.2+2419	&	J11503+2417	&	11:50:19.2	&	+24:17:53.852	&	1	&	BLL	&	0.2	&	1	&	0.3	&	7.60E+10	&	17	&	7	&	745.7	\\
1FGL J1151.6+5857	&	J11514+5859	&	11:51:24.7	&	+58:59:17.552	&	1	&	BLL	&	\nodata	&	0.7	&	0.3	&	3.08E+09	&	35	&	10	&	102.2	\\
1FGL J1152.1+6027	&	J11540+6022	&	11:54:04.5	&	+60:22:20.785	&	0.73	&	\nodata	&	\nodata	&	0.6	&	0.3	&	1.71E+11	&	\nodata	&	\nodata	&	151.6	\\
1FGL J1202.9+6032	&	J12030+6031	&	12:03:03.5	&	+60:31:19.129	&	1	&	RG	&	0.065	&	1.1	&	0.3	&	7.22E+09	&	21	&	12	&	191.4	\\
1FGL J1209.3+5444	&	J12089+5441	&	12:08:54.3	&	+54:41:58.190	&	0.97	&	FSRQ	&	1.344	&	0.6	&	0.2	&	1.17E+11	&	\nodata	&	\nodata	&	262	\\
1FGL J1209.4+4119	&	J12093+4119	&	12:09:22.8	&	+41:19:41.360	&	1	&	BLL	&	\nodata	&	0.5	&	0.2	&	2.99E+10	&	\nodata	&	\nodata	&	485.1	\\
1FGL J1209.7+1806	&	J12098+1810	&	12:09:51.8	&	+18:10:06.796	&	0.97	&	FSRQ	&	0.845	&	0.8	&	0.2	&	4.41E+10	&	\nodata	&	\nodata	&	136.3	\\
1FGL J1217.7+3007	&	J12178+3007	&	12:17:52.1	&	+30:07:00.625	&	1	&	BLL	&	0.13	&	6.7	&	0.6	&	5.00E+10	&	19	&	0	&	334.9	\\
1FGL J1220.2+3432	&	J12201+3431	&	12:20:08.3	&	+34:31:21.711	&	1	&	BLL	&	\nodata	&	0.6	&	0.2	&	3.05E+10	&	\nodata	&	\nodata	&	314.1	\\
1FGL J1221.5+2814	&	J12215+2813	&	12:21:31.7	&	+28:13:58.497	&	1	&	BLL	&	0.102	&	6.9	&	0.5	&	2.30E+10	&	27	&	-12	&	893.1	\\
1FGL J1225.8+4336	&	J12248+4335	&	12:24:51.5	&	+43:35:19.276	&	0.94	&	\nodata	&	\nodata	&	0.8	&	0	&	1.80E+10	&	\nodata	&	\nodata	&	220.8	\\
1FGL J1225.8+4336	&	J12269+4340	&	12:26:57.9	&	+43:40:58.438	&	0.87	&	FSRQ	&	2.002	&	0.8	&	0	&	1.07E+10	&	\nodata	&	\nodata	&	145.1	\\
1FGL J1228.2+4855	&	J12288+4858	&	12:28:51.8	&	+48:58:01.293	&	0.91	&	FSRQ	&	1.722	&	1.3	&	0.3	&	1.89E+12	&	17	&	-6	&	265.5	\\
1FGL J1230.4+2520	&	J12302+2518	&	12:30:14.1	&	+25:18:07.145	&	0.99	&	BLL	&	0.135	&	1.4	&	0.3	&	1.96E+10	&	30	&	-9	&	326.4	\\
1FGL J1248.2+5820	&	J12483+5820	&	12:48:18.8	&	+58:20:28.725	&	1	&	BLL	&	\nodata	&	4.5	&	0.4	&	1.54E+10	&	43	&	15	&	324.4	\\
1FGL J1253.0+5301	&	J12531+5301	&	12:53:11.9	&	+53:01:11.741	&	1	&	BLL	&	\nodata	&	3	&	0.4	&	3.05E+10	&	14	&	5	&	366	\\
1FGL J1258.3+3227	&	J12579+3229	&	12:57:57.2	&	+32:29:29.321	&	0.98	&	FSRQ	&	0.806	&	0.8	&	0	&	2.26E+11	&	\nodata	&	\nodata	&	477.1	\\
1FGL J1303.0+2433	&	J13030+2433	&	13:03:03.2	&	+24:33:55.684	&	1	&	BLL	&	\nodata	&	3.5	&	0.4	&	7.16E+10	&	\nodata	&	\nodata	&	135.7	\\
1FGL J1308.5+3550	&	J13083+3546	&	13:08:23.7	&	+35:46:37.160	&	0.99	&	FSRQ	&	1.055	&	2.2	&	0.3	&	2.15E+11	&	60	&	-4	&	527	\\
1FGL J1312.4+4827	&	J13127+4828	&	13:12:43.4	&	+48:28:30.928	&	0.99	&	FSRQ	&	0.501	&	1.4	&	0.3	&	2.24E+10	&	\nodata	&	\nodata	&	250.3	\\
1FGL J1314.7+2346	&	J13147+2348	&	13:14:43.8	&	+23:48:26.701	&	1	&	BLL	&	\nodata	&	1.4	&	0.3	&	6.62E+09	&	\nodata	&	\nodata	&	158.2	\\
1FGL J1317.8+3425	&	J13176+3425	&	13:17:36.5	&	+34:25:15.921	&	0.98	&	FSRQ	&	1.05	&	0.5	&	0.2	&	1.54E+10	&	\nodata	&	\nodata	&	541.3	\\
1FGL J1321.1+2214	&	J13211+2216	&	13:21:11.2	&	+22:16:12.098	&	0.99	&	FSRQ	&	0.943	&	1.1	&	0.3	&	1.83E+10	&	\nodata	&	\nodata	&	323.6	\\
1FGL J1331.0+5202	&	J13307+5202	&	13:30:42.6	&	+52:02:15.448	&	0.99	&	RG	&	0.688	&	0.5	&	0	&	5.43E+10	&	\nodata	&	\nodata	&	171.9	\\
1FGL J1332.9+4728	&	J13327+4722	&	13:32:45.2	&	+47:22:22.653	&	0.96	&	FSRQ	&	0.669	&	0.5	&	0.2	&	1.02E+11	&	\nodata	&	\nodata	&	457.5	\\
1FGL J1333.2+5056	&	J13338+5057	&	13:33:53.8	&	+50:57:35.914	&	0.93	&	\nodata	&	\nodata	&	1.4	&	0.3	&	6.24E+09	&	\nodata	&	\nodata	&	107.6	\\
1FGL J1345.4+4453	&	J13455+4452	&	13:45:33.2	&	+44:52:59.581	&	0.99	&	FSRQ	&	2.534	&	1.6	&	0.3	&	2.01E+11	&	\nodata	&	\nodata	&	197.5	\\
1FGL J1351.0+3035	&	J13508+3034	&	13:50:52.7	&	+30:34:53.582	&	1	&	FSRQ	&	0.714	&	0.6	&	0.2	&	3.31E+10	&	\nodata	&	\nodata	&	719.9	\\
1FGL J1359.1+5539	&	J13590+5544	&	13:59:05.7	&	+55:44:29.362	&	0.98	&	FSRQ	&	1.014	&	0.9	&	0.2	&	9.56E+10	&	\nodata	&	\nodata	&	143.8	\\
1FGL J1421.0+5421	&	J14197+5423	&	14:19:46.6	&	+54:23:14.783	&	0.71	&	BLL	&	0.152	&	0.9	&	0.2	&	5.81E+10	&	20	&	4	&	2247.3	\\
1FGL J1426.9+2347	&	J14270+2348	&	14:27:00.4	&	+23:48:00.045	&	1	&	BLL	&	\nodata	&	10.2	&	0.6	&	8.28E+09	&	73	&	-84	&	241.6	\\
1FGL J1433.9+4204	&	J14340+4203	&	14:34:05.7	&	+42:03:16.010	&	1	&	FSRQ	&	1.24	&	0.7	&	0.2	&	4.29E+10	&	\nodata	&	\nodata	&	286.8	\\
1FGL J1436.9+2314	&	J14366+2321	&	14:36:41.0	&	+23:21:03.297	&	0.96	&	FSRQ	&	1.545	&	0.5	&	0.2	&	1.40E+11	&	\nodata	&	\nodata	&	650.7	\\
1FGL J1438.7+3711	&	J14388+3710	&	14:38:53.6	&	+37:10:35.408	&	0.99	&	FSRQ	&	2.401	&	0.7	&	0.2	&	1.45E+09	&	35	&	22	&	315.4	\\
1FGL J1451.0+5204	&	J14509+5201	&	14:51:00.0	&	+52:01:11.700	&	0.99	&	BLL	&	\nodata	&	0.9	&	0.3	&	2.86E+09	&	\nodata	&	\nodata	&	86.9	\\
1FGL J1454.6+5125	&	J14544+5124	&	14:54:27.1	&	+51:24:33.734	&	1	&	BLL	&	\nodata	&	1.1	&	0.3	&	6.16E+09	&	\nodata	&	\nodata	&	96.3	\\
1FGL J1503.3+4759	&	J15037+4759	&	15:03:48.0	&	+47:59:31.024	&	0.96	&	BLL	&	\nodata	&	0.6	&	0.2	&	4.09E+10	&	\nodata	&	\nodata	&	94.1	\\
1FGL J1505.8+3725	&	J15061+3730	&	15:06:09.5	&	+37:30:51.128	&	0.98	&	FSRQ	&	0.674	&	0.7	&	0.2	&	1.40E+11	&	18	&	4	&	777.3	\\
1FGL J1516.9+1928	&	J15169+1932	&	15:16:56.8	&	+19:32:13.010	&	1	&	BLL	&	\nodata	&	0.6	&	0.3	&	3.94E+10	&	\nodata	&	\nodata	&	609.3	\\
1FGL J1522.1+3143	&	J15221+3144	&	15:22:10.0	&	+31:44:14.427	&	1	&	FSRQ	&	1.487	&	15.9	&	0.8	&	1.45E+11	&	\nodata	&	\nodata	&	493.8	\\
1FGL J1539.7+2747	&	J15396+2744	&	15:39:39.1	&	+27:44:38.288	&	1	&	FSRQ	&	2.19	&	1	&	0.2	&	1.29E+11	&	\nodata	&	\nodata	&	291.5	\\
1FGL J1542.9+6129	&	J15429+6129	&	15:42:56.9	&	+61:29:55.358	&	1	&	BLL	&	\nodata	&	5.2	&	0.4	&	5.94E+10	&	\nodata	&	\nodata	&	144	\\
1FGL J1558.9+5627	&	J15588+5625	&	15:58:48.3	&	+56:25:14.108	&	1	&	BLL	&	0.3	&	1.8	&	0.3	&	1.30E+10	&	\nodata	&	\nodata	&	169.7	\\
1FGL J1604.3+5710	&	J16046+5714	&	16:04:37.4	&	+57:14:36.668	&	0.98	&	FSRQ	&	0.72	&	1.2	&	0.3	&	1.61E+10	&	\nodata	&	\nodata	&	485.3	\\
1FGL J1607.1+1552	&	J16071+1551	&	16:07:06.4	&	+15:51:34.495	&	1	&	RG	&	0.496	&	2.4	&	0.4	&	2.80E+10	&	13	&	-1	&	223.5	\\
1FGL J1616.1+4637	&	J16160+4632	&	16:16:03.8	&	+46:32:25.239	&	0.96	&	FSRQ	&	0.95	&	0.6	&	0.2	&	2.77E+11	&	\nodata	&	\nodata	&	125.5	\\
1FGL J1635.0+3808	&	J16352+3808	&	16:35:15.5	&	+38:08:04.497	&	0.99	&	FSRQ	&	1.814	&	6.8	&	0.5	&	2.47E+10	&	\nodata	&	\nodata	&	2423.3	\\
1FGL J1637.9+4707	&	J16377+4717	&	16:37:45.1	&	+47:17:33.822	&	0.91	&	FSRQ	&	0.74	&	1	&	0	&	6.85E+10	&	\nodata	&	\nodata	&	815.6	\\
1FGL J1637.9+4707	&	J16399+4705	&	16:39:56.0	&	+47:05:23.575	&	0.51	&	FSRQ	&	0.86	&	1	&	0	&	2.28E+09	&	\nodata	&	\nodata	&	108.7	\\
1FGL J1647.4+4948	&	J16475+4950	&	16:47:34.9	&	+49:50:00.586	&	0.99	&	RG	&	0.047	&	1.1	&	0.3	&	2.47E+10	&	\nodata	&	\nodata	&	189	\\
1FGL J1653.9+3945	&	J16538+3945	&	16:53:51.8	&	+39:45:31.490	&	0.99	&	BLL	&	0.034	&	8.3	&	0.6	&	8.10E+10	&	17	&	12	&	1153.2	\\
1FGL J1656.9+6017	&	J16568+6012	&	16:56:48.2	&	+60:12:16.455	&	0.96	&	FSRQ	&	0.623	&	0.8	&	0.2	&	2.51E+11	&	\nodata	&	\nodata	&	226.5	\\
1FGL J1709.6+4320	&	J17096+4318	&	17:09:41.1	&	+43:18:44.547	&	1	&	FSRQ	&	1.027	&	1.5	&	0.3	&	9.36E+10	&	40	&	10	&	197.2	\\
1FGL J1724.0+4002	&	J17240+4004	&	17:24:05.4	&	+40:04:36.457	&	1	&	RG	&	1.049	&	2.9	&	0.4	&	6.08E+10	&	9	&	7	&	296.1	\\
1FGL J1727.3+4525	&	J17274+4530	&	17:27:27.6	&	+45:30:39.743	&	0.97	&	FSRQ	&	0.714	&	1.2	&	0.3	&	1.95E+11	&	\nodata	&	\nodata	&	1360.3	\\
1FGL J1727.9+5010	&	J17283+5013	&	17:28:18.6	&	+50:13:10.480	&	0.98	&	BLL	&	0.055	&	0.9	&	0.2	&	4.82E+09	&	18	&	22	&	161.8	\\
1FGL J1734.4+3859	&	J17343+3857	&	17:34:20.6	&	+38:57:51.446	&	1	&	FSRQ	&	0.976	&	6	&	0.5	&	1.52E+11	&	\nodata	&	\nodata	&	1101.3	\\
1FGL J1740.0+5209	&	J17406+5211	&	17:40:37.0	&	+52:11:43.413	&	0.78	&	FSRQ	&	1.379	&	3.4	&	0.4	&	5.75E+10	&	\nodata	&	\nodata	&	1358.5	\\
1FGL J1742.1+5947	&	J17425+5945	&	17:42:32.0	&	+59:45:06.729	&	0.97	&	BLL	&	\nodata	&	0.6	&	0.2	&	3.86E+10	&	\nodata	&	\nodata	&	116.2	\\
1FGL J1749.0+4323	&	J17490+4321	&	17:49:00.4	&	+43:21:51.287	&	1	&	BLL	&	\nodata	&	2.3	&	0.3	&	6.17E+10	&	21	&	-10	&	284.9	\\
\enddata
\tablecomments{Col.\ (1): 1FGL source name.  
Col.\ (2): VIPS source name.
Col.\ (3): Right Ascension (J2000).
Col.\ (4): Declination (J2000).
Col.\ (5): Probability that the LAT source is associated with the VIPS source.
Col.\ (6): Source classification. 
Col.\ (7): Redshift.
Col.\ (8): $\gamma$-ray flux in units of $10^{-9}$ photons cm$^{-2}$ s$^{-1}$ for 1-100 GeV.  NOTE: This is not the full 100 MeV to 100 GeV flux.
Col.\ (9): Error in the $\gamma$-ray flux in the same units as the flux.
Col.\ (10): Core Brightness Temperature as measured by automated program (Helmboldt et al 2007).
Col.\ (11): Opening Angle in degrees
Col.\ (12): Change in jet position angle in degrees.
Col.\ (13): CLASS flux density at 8.5 GHz in mJy.
}
\label{datatable}
\end{deluxetable}

\textwidth = 7.0truein
\textheight = 10.0truein
\begin{deluxetable}{ccrrrrrr}
\tablecolumns{8}
\tabletypesize{\scriptsize}
\tablewidth{0pt}
\tablecaption{Source Classifications}
\tablehead{
\colhead{LAT/non-LAT} & \colhead{Opt Type}	&	\colhead{LJET}	&	\colhead{SJET}	&	\colhead{PS}	&	\colhead{CPLX}	&	\colhead{CSO}	&	\colhead{Unidentified} \\}
\startdata
LAT-detected &  &  &  &  &  &  &  \\
 & BL Lacs & 30 (75\%) & 4 (10\%) & 4 (10\%) & 2 (5\%) & \nodata & \nodata \\
 & FSRQs & 22 (44\%) & 11 (22\%) & 16 (32\%) & 1 (2\%) & \nodata & \nodata \\
 & RG/Other & 8 (67\%) & 1 (8\%) & 3 (25\%) & \nodata & \nodata & \nodata \\
non-LAT-detected &  &  &  &  &  &  &  \\
 & BL Lacs & 11 (46\%) & 7 (29\%) & 6 (25\%) & \nodata & \nodata & \nodata \\
 & FSRQs & 188 (39\%) & 121 (25\%) & 136 (28\%) & 2 ($\sim$1\%) & 30 (6\%) & 2 ($\sim$1\%) \\
 & RG/Other & 214 (42\%) & 98 (19\%) & 111 (21\%) & 11 (2\%) & 71 (14\%) & 10 (2\%) \\
\enddata
\tablecomments{LJET = long jet, SJET = short jet, PS = point source, CPLX = complex, CSO = compact symmetric object candidate
}
\label{morphtable}
\end{deluxetable}

\clearpage
\textwidth = 7.2truein

\clearpage

\clearpage

\clearpage

\clearpage

\end{document}